\documentclass[prb,aps,superscriptaddress,twocolumn]{revtex4}
\usepackage{graphicx}
\usepackage{dcolumn}
\usepackage{bm}

\usepackage{color} 

\usepackage{ulem}

\providecommand{\U}[1]{\protect\rule{.1in}{.1in}}

\begin{document}


\title{Selection of direction of the ordered moments in Na$_2$IrO$_3$ and $\alpha-$RuCl$_3$}

\author{Yuriy Sizyuk}
\affiliation{School of Physics and Astronomy, University of Minnesota, Minneapolis,
MN 55116, USA}

\author{Peter W\"{o}lfle}
    \affiliation{Institute for Condensed Matter Theory and Institute for Nanotechnology, Karlsruhe Institute of Technology, D-76128 Karlsruhe, Germany}

\author{Natalia  B. Perkins}

\affiliation{School of Physics and Astronomy, University of Minnesota, Minneapolis,
MN 55116, USA}

\begin{abstract}
The magnetic orders in Na$_2$IrO$_3$ and $\alpha-$RuCl$_3$, honeycomb systems with strong spin-orbit coupling and correlations, have been recently described by models with the dominant Kitaev interactions. In this work we discuss how the orientation of the magnetic order parameter is selected in this class of models. We show that while the order-by-disorder mechanism in the models with solely Kitaev anisotropies always select cubic axes as easy axes for magnetic ordering, the additional effect of other small bond-dependent anisotropies, such as, e.g., $\Gamma$-terms, lead to a deviation of the order parameter from the cubic directions. We show that both the zigzag ground state and the face-diagonal orientation of the magnetic moments in Na$_2$IrO$_3$ can be obtained within the $J_1-K_1-J_2-K_2-J_3$ model in the presence of perturbatively small $\Gamma$-terms. We also show that the zigzag phase found in the nearest neighbor Kitaev-Heisenberg model, relevant for $\alpha-$RuCl$_3$, has some stability against the $\Gamma$-term.
\end{abstract}

\maketitle

\section{Introduction}
 The long-standing quest for a solid state realization  of the Kitaev honeycomb model\cite{kitaev06}
has triggered much of the experimental and theoretical interest in
4d and 5d compounds with two- and three-dimensional tri-coordinated lattices, in which
  the interplay of the strong spin-orbit coupling (SOC)
and electronic  correlations leads to the  dominance of the strongly anisotropic  Kitaev-like interactions.\cite{jackeli09} A lot of experimental  effort has been focused on iridium oxides belonging
 to  the A$_2$IrO$_3$  family\cite{singh10,singh12,liu11,ye12,choi12,gretarsson13,chun15,modic14, biffin14-1,biffin14-2,tomo15} 
 and, more recently, to  $\alpha-$RuCl$_3$.\cite{polini96,plumb14,sears15,majumber15,johnson15,banerjee15,cao16}
 
 The Kitaev honeycomb  model  belongs to the class of the compass models. It is intrinsically frustrated due to  the bond-depended nature of the interactions. In the quantum case, this frustration leads to the appearance of the non-trivial quantum spin liquid (QSL)  phase with fractionalized
excitations, dubbed Kitaev QSL.\cite{kitaev06}
 Kitaev QSL is not a unique example of non-trivial ground states  of the compass models, \cite{kha05,nussinov15} however, it is probably the only one which allows  an exact analytic solution.

  In honeycomb  iridates and ruthenates, the magnetic degree of
freedom described by an  effective magnetic moment  $J_{\rm eff}=1/2$, arises  in the presence of strong SOC from electrons occupying  $t_{2g}$-manifold of states  of  Ir$^{4+}$  and  Ru$^{3+}$ ions.
In  A$_2$IrO$_3$ compounds, edge-shared IrO$_6$  octahedra provide  90$^{\circ}$ paths for the dominant  nearest neighbor Kitaev  coupling between  iridium  magnetic moments. A similar situation takes place  in  the  isostructural  layered honeycomb material $\alpha-$RuCl$_3$ and three-dimensional harmonic honeycomb iridates, $\beta-$Li$_2$IrO$_3$ and $\gamma-$Li$_2$IrO$_3$.\cite{modic14, biffin14-1,biffin14-2,tomo15}
  It is believed that  the sign of  the Kitaev interaction  may be  either antiferro-  (AF)  or ferromagnetic (FM) depending on the compound.\cite{jackeli10,jackeli13,katukuri14,sizyuk14,rachel14,ioannis15,kee15}
 Isotropic  Heisenberg  couplings  are also present in these compounds due to the octachedra edge sharing geometry and direct overlap of $5d-$  or $4d-$orbitals which,  due to their extended nature, often reach beyond  nearest neighbors.
 Further anisotropies, such as the isotropic off-diagonal $\Gamma$ interactions, can also be present, mainly as a result of crystal field distortions. \cite{sizyuk14, rau14,chaloupka15}
The competition between all these couplings leads  to a rich variety of  experimentally observed magnetic structures.\cite{liu11,ye12,choi12,gretarsson13,chun15,modic14,biffin14-1,biffin14-2,tomo15}  
 
  Here we discuss in detail the  models  and the mechanisms which lead to the stabilization of  magnetic ordering in  two compounds: Na$_2$IrO$_3$ and $\alpha-$RuCl$_3$.
  Several experiments have shown  that the low-temperature phase of Na$_2$IrO$_3$ has collinear zigzag long-range magnetic order.\cite{singh10,singh12,liu11,ye12,choi12,gretarsson13,chun15}
 In addition, recent diffuse magnetic x-ray scattering data have determined the spin orientation in this zigzag state  and showed that it is along  the 44.3$^\circ$ direction  with respect to  the $a$ axis, which corresponds to approximately half way in between the cubic $x$ and $y$ axes.\cite{chun15} 
 Both of these findings are in  disagreement with the original Kitaev-Heisenberg model,\cite{jackeli10,jackeli13} which  predicts the zigzag phase only for the antiferromagnetic nearest neighbor Kitaev interaction with the magnetic moments along the cubic axes, while the Kitaev interaction in Na$_2$IrO$_3$ is ferromagnetic.\cite{katukuri14} This shows that  one needs to extend the nearest neighbor model by including some additional  interactions in order to explain these experimental observations.

$\alpha-$RuCl$_3$ also shows  collinear antiferromagnetic zigzag  ground state.\cite{sears15,johnson15,banerjee15,cao16}   Recent  X-ray  and neutron scattering diffraction data\cite{banerjee15,cao16} indicate that  the best fit to the collinear structure  is  obtained for the antiferromagnetic  nearest neighbor   Kitaev interaction and when the spin direction  points 35$^\circ$ out of the $ab$-plane, i.e. along one of the cubic directions. 
 This suggests that  the microscopic origin of the zigzag ground state in $\alpha-$RuCl$_3$
might be quite different from the one in Na$_2$IrO$_3$, and  that it can be described by the nearest neighbor Kitaev-Heisenberg model.\cite{jackeli13}

  In both cases,  the  available experimental data  provides an important check of the validity of any  model proposed to describe the  magnetic properties  of Na$_2$IrO$_3$   and $\alpha-$RuCl$_3$, as it should correctly predict not only the type of the magnetic order but also  its orientation in space.

 In this work   we consider  two models, the nearest neighbor Kitaev-Heisenberg  model\cite{jackeli09,jackeli10,jackeli13} 
and its more complicated counterpart, dubbed $J_1-K_1-J_2-K_2-J_3$ model,\cite{sizyuk14}  and
  study how  the preferred directions of the mean
field order parameter  are selected in these models.  The formal procedure  which we will be using here is based on the derivation of the fluctuational part of the free energy by integrating out the leading thermal fluctuations, and by determining which orientations of the order parameter correspond to the free energy minimum.  This approach  is based on the Hubbard-Stratonovich  transformation and  was outlined  in Refs.\cite{sizyuk15,peter15}
to which we refer the reader for technical details.  In both models, the thermal fluctuations select the cubic axes as the preferred directions for spins, which describes the experimental situation in $\alpha-$RuCl$_3$ but not in Na$_2$IrO$_3$.

 We have  also checked that  in both models the  quantum fluctuations (taken into account either  using the quantum version of Hubbard-Stratonovich approach or within the semiclassical spin-wave approach) lift the accidental degeneracy of the  classical solution and also select the cubic axes as the preferred directions for spins. We did not present these calculations here as they bring no new results compared with more simple analysis of  thermal fluctuations.  
 
 The important point which we stress in our paper is that the selection of correct  "diagonal" direction of the spins   observed in Na$_2$IrO$_3$  might happen already on the mean-field level by inclusion of small off-diagonal positive interaction $\Gamma$ as soon as it is larger than the  energy gain  of order $1/S$  due to  the quantum fluctuations.

This paper is organized as follows.
 In Sec. II   we  study  the order by disorder  mechanism of the selection  of the direction of the order parameter  in the  nearest neighbor Kitaev-Heisenberg model on the honeycomb lattice. In Sec. III we extend our consideration to  the $J_1-K_1-J_2-K_2-J_3$ model.
 In Sec. IV, we  discuss the role of the off-diagonal $\Gamma$-term and study  the selection of the direction of the magnetic order in Na$_2$IrO$_3$ and in  $\alpha-$RuCl$_3$.  
We summarize our conclusions in Sec. V. 
Appendix A discusses in detail   the degeneracy of the classical manifold of the Kitaev-Heisenberg model.
Appendices B and C contain  some technical details.

\begin{figure}
\includegraphics[width=0.65\columnwidth]{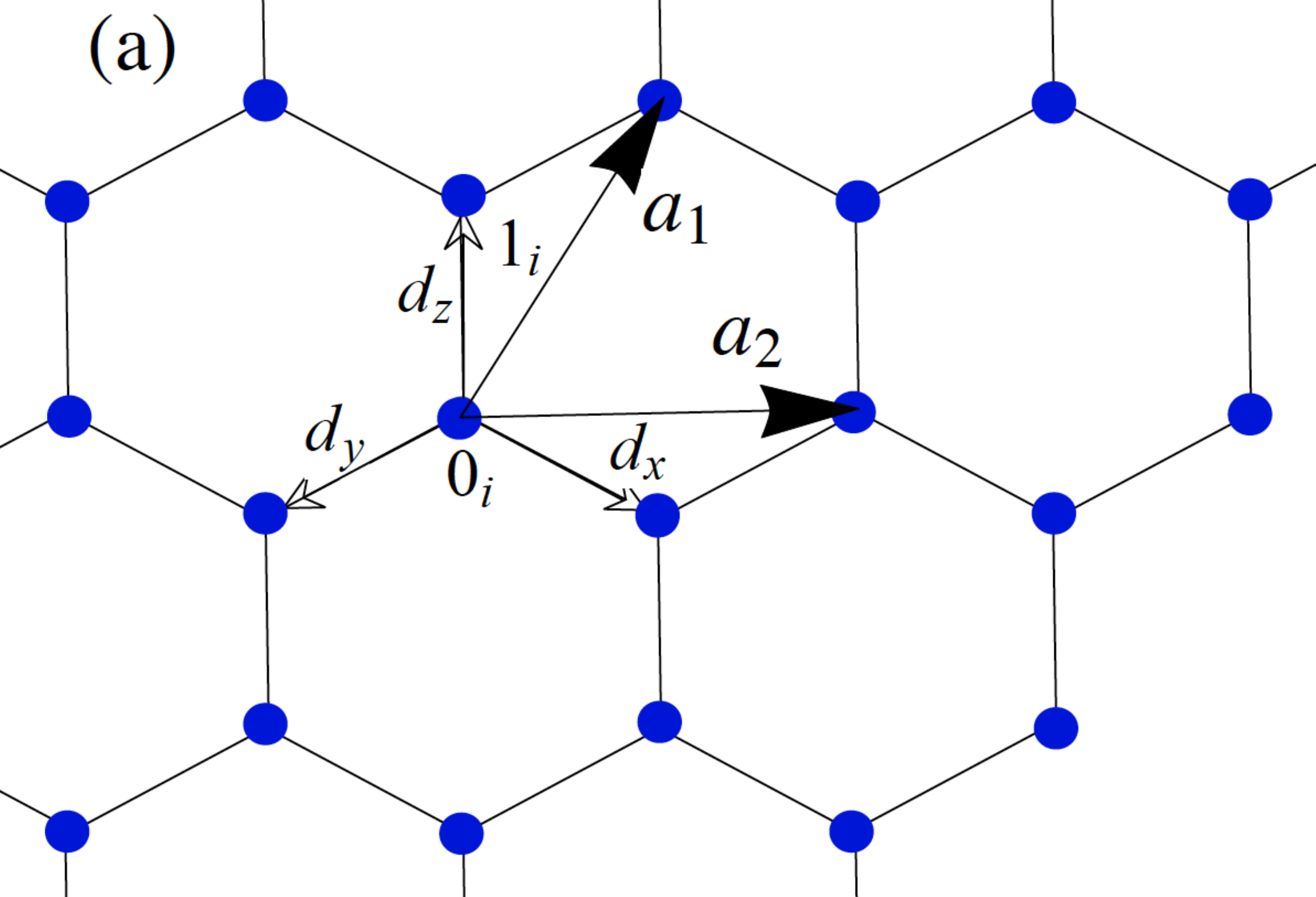}
\includegraphics[width=0.65\columnwidth]{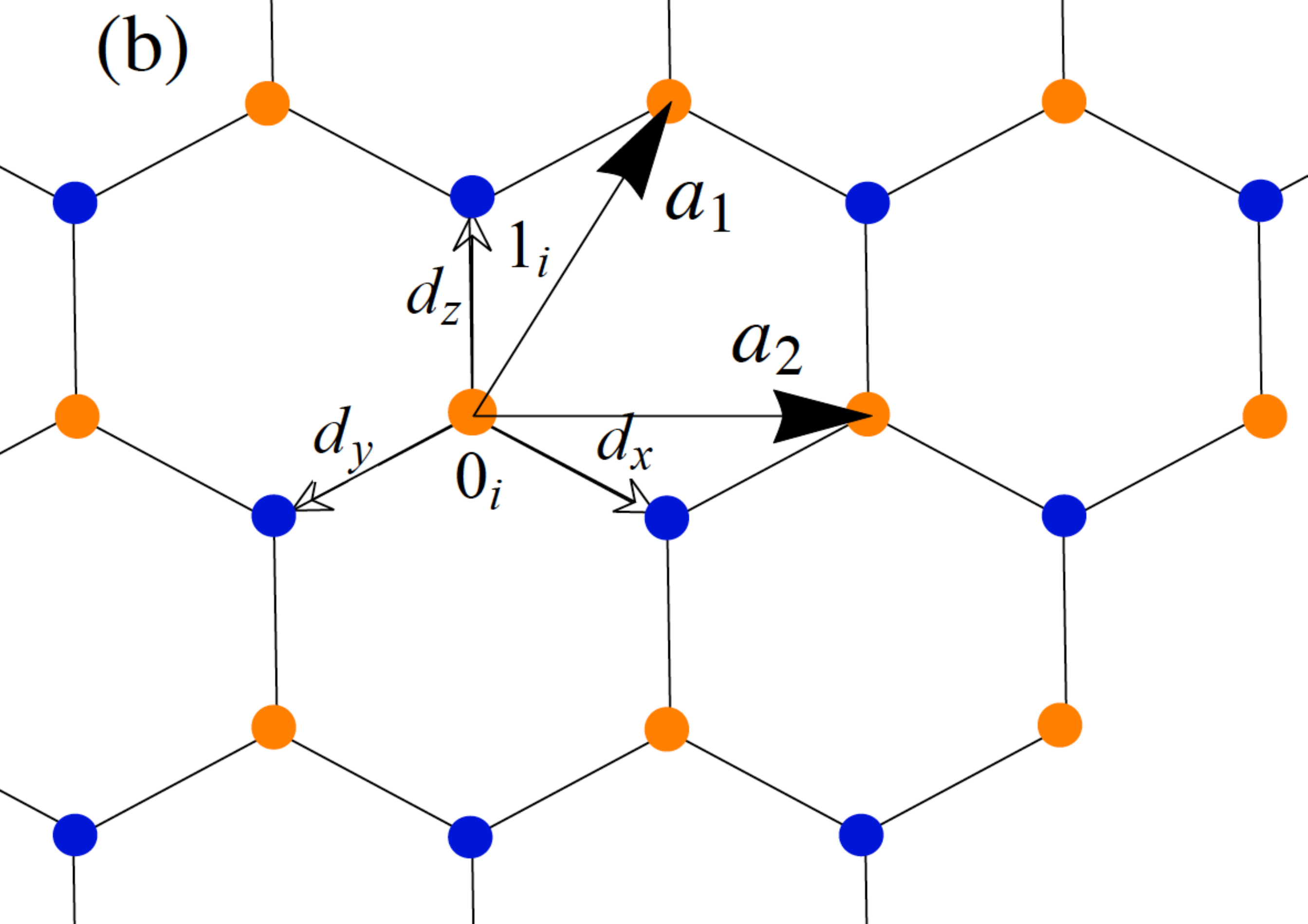}
\includegraphics[width=0.65\columnwidth]{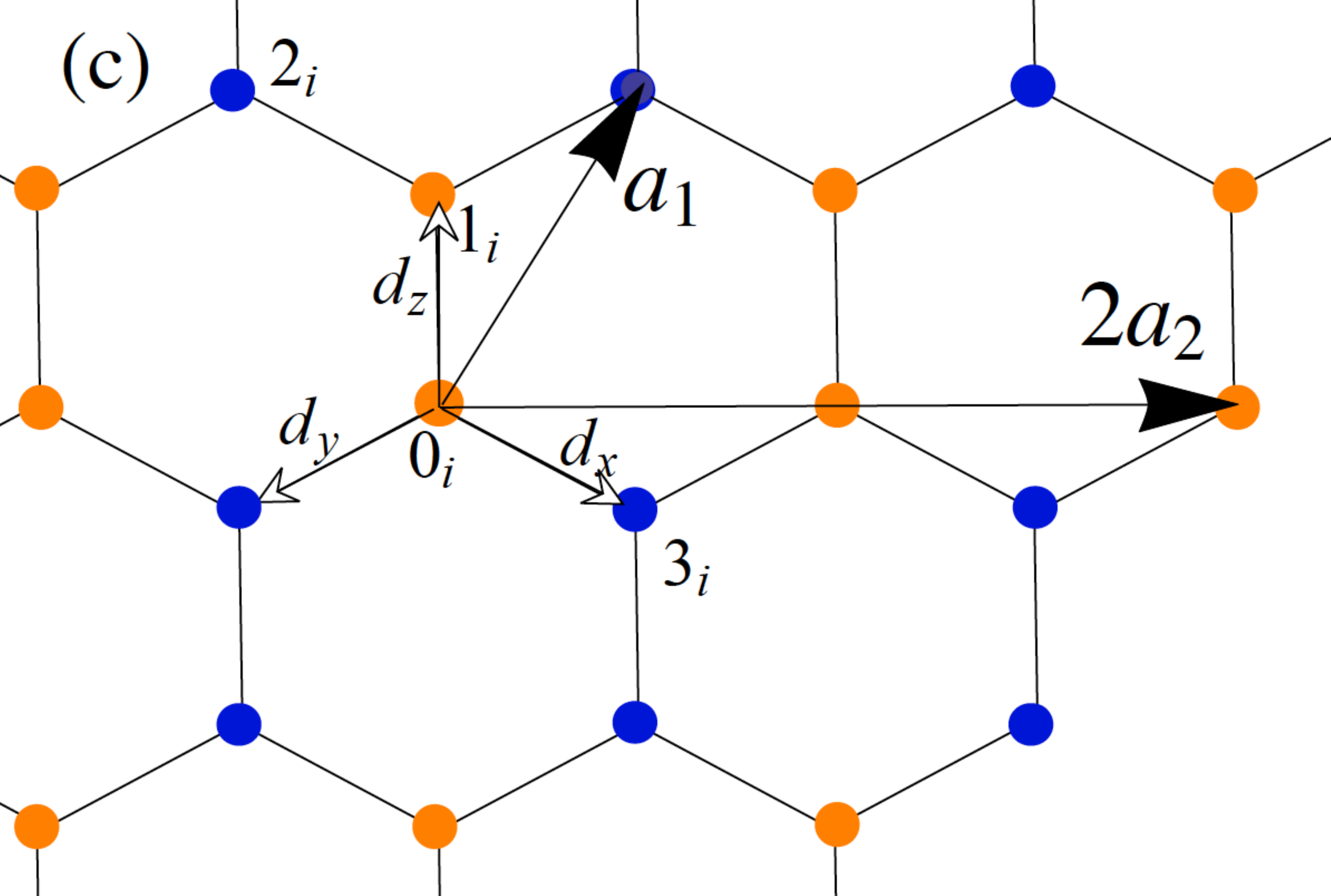}
\includegraphics[width=0.65\columnwidth]{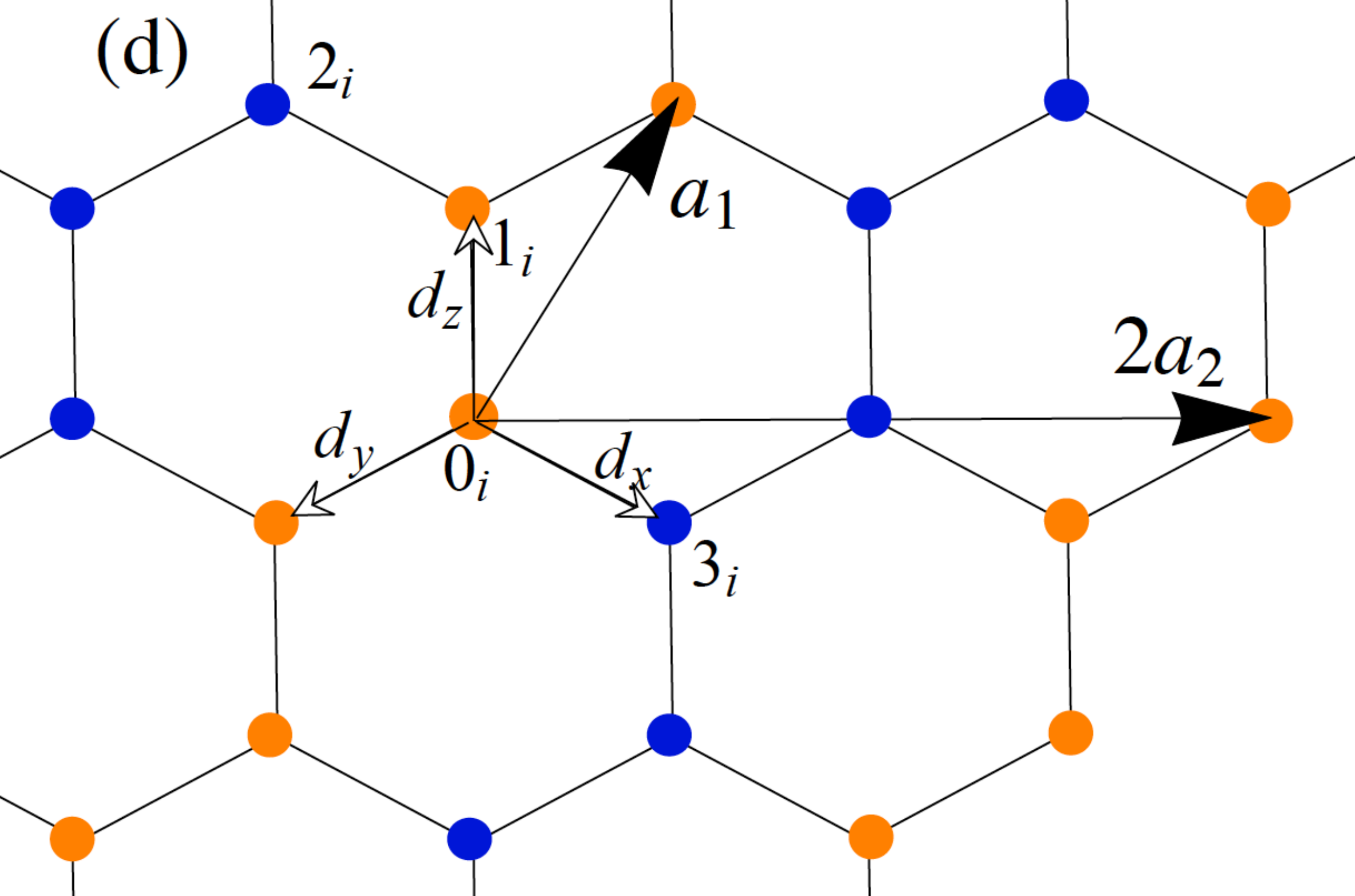}
\caption{
Four possible magnetic configurations:
(a) FM ordering;
(b) AF N\'{e}el order;
(c) AF stripy  order;
(d)  AF zigzag order.
Red and blue circles correspond to up and down spins.
Here $\bm{a}_1\!=\!(\!\frac{\sqrt{3}}{2}\hat{x}+\frac{3}{2}\hat{y}\!) $ and 
$\bm{a}_2\!=\!\sqrt{3}\hat{x}$   are two primitive translations.  
 The bond vectors are $\bm{d}_x\!=\!(\!\frac{\sqrt{3}}{2}\hat{x}-\frac{1}{2}\hat{y}\!) $, 
 $\bm{d}_y\!=\!(\!-\frac{\sqrt{3}}{2}\hat{x}-\frac{1}{2}\hat{y}\!) $ and 
$\bm{d}_z\!=\!\hat{y}$.   }
\label{fig:orders}
\end{figure}

\begin{figure}
\includegraphics[width=0.95\columnwidth]{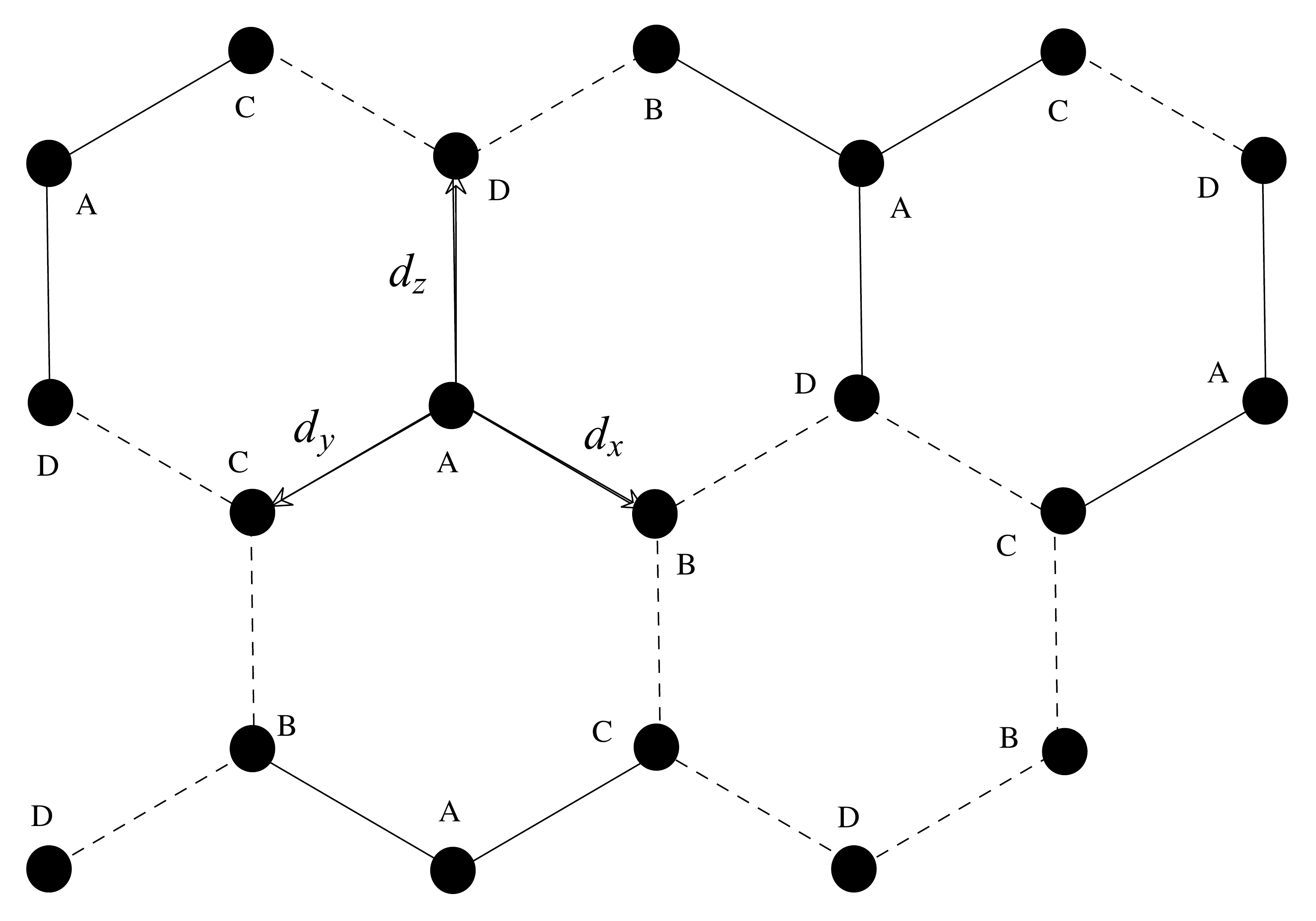}
\caption{A, B, C and D designate the four sublattices of the Klein transformation.
  Solid and dashed bonds shows the change of the sign of the  $\Gamma$ interaction in the  four-sublattice transformation:   $\Gamma$  picks up a negative sign on the solid bonds but keeps the sign from unrotated reference frame on the dashed bonds.  }
\label{fig:orders1}
\end{figure}

\section{Order by disorder in the extended nearest neighbor Kitaev-Heisenberg model}
 
 The  Kitaev-Heisenberg model  on the honeycomb lattice reads \cite{jackeli13}
\begin{eqnarray}\label{KH}
H&=&\sum_{\langle ij \rangle_\alpha}\sum_{\gamma}J^{\alpha\gamma}S_{0,i}^{\gamma
}S_{1,j}^{\gamma},
\end{eqnarray}
where    $J^{\alpha\gamma}=J+K\delta
_{\alpha,\gamma}$ is the interaction between  $\gamma$-component of the  pseudospin $S^{\gamma}_ {\nu,j}=1/2$, on sublattices $\nu=0,1$. Hereafter, we
call these pseudospins  simply spins.  $J$ and $K$ correspond to the Heisenberg and Kitaev interactions, which in the extended model can be both AF and FM. 
$\gamma=x,y,z$ denote  the  spin components in the global  reference frame.

The classical  phase diagram of the model (1) contains four magnetic phases:\cite{jackeli13,price13}
 the ferromagnetic phase  (Fig. 1 (a)), the  N\'{e}el antiferromagnet
 (Fig. 1(b)), the  stripy antiferromagnet  (Fig. 1 (c)) and the zigzag antiferromagnet (Fig. 1 (d)).   
 The latter two magnetic states have a four sublattice structure.
 
 All these phases have macroscopic classical degeneracy.
 While the classical degeneracy of the  simple FM state  and  of  the AF N\'{e}el state comes straightforwardly from the infinite number of degenerate collinear states, the macroscopic  degeneracy of  the AF stripy  and zigzag phases  is more complex, and  the 
 degenerate ground state manifold consists of six collinear states  and  a  set of  non-collinear multi-$Q$ states. 
 In Appendix A we discuss this question in detail and show that using the  four-sublattice Klein transformation   for the stripy and the zigzag AF states,\cite{jackeli10,kimchi14,chaloupka15}  
the nature of the classical degeneracy of all four  magnetically ordered states can be understood  in a similar way. Importantly, in all cases,  the classical degeneracy is accidental and is 
removed by the order by disorder mechanism which selects a set of collinear states, each  with a   particular direction of the order parameter. 

 Following  Chaloupka {\it et al},\cite{jackeli10}  we introduce four  auxiliary  sublattices A, B, C and D (see Fig.2), fix the direction of the spins on the  sublattice  A and rotate the spins on the subllatices B, C, and D such that
 the   component of spin corresponding  to the bond direction ($x$ for B, $y$ for C and $z$ for D) stays the same but two other spin components change sign. This results in the transformed Hamiltonian with the same form as (\ref{KH}) but with transformed couplings.

 Here we consider the Kitaev-Heisenberg model  in the full parameter space.  For the parameters of the model for which either stripy or zigzag  are the ground states, we perform  four-sublattice transformation and treat the model (\ref{KH}) in the rotated  basis, in which the stripy order maps to the FM and the zigzag order maps to  the simple  two-sublattice AFM  N\'{e}el  state.

 Next, using a Hubbard-Stratonovich transformation  of the partition function,\cite{sizyuk15, peter15}  we discuss  how the preferred directions  of the order parameter in all these phases  are selected by thermal    below the  ordering temperature.

The partition function of the system of classical spins is given by the  integral over the Boltzmann weights of the configurations
\begin{eqnarray}\label{Z-clas}
&&Z=\int\int\left[d\mathbf{S}_{0,i}][d\mathbf{S}_{1,j}\right]\delta(\mathbf{S}_{0,i}^{2}-1)\delta(\mathbf{S}_{1,j}^{2}-1)\nonumber\\&&
\exp\left[-\beta\sum_{\langle ij \rangle_\alpha}\sum_{\gamma}J^{\alpha\gamma}S_{0,i}^{\gamma
}S_{1,j}^{\gamma}\right],
\end{eqnarray}
where $\mathbf{S}_{0,j}$ and $\mathbf{S}_{1,j}$ are classical spins on sublattices $0$ and $1$, and  $\beta=1/(k_BT)$ is the inverse temperature.
Similarly in the case of a quantum system the partition function is a trace of the Boltzman weights over the spin operators,
$
Z=Tr\left[\exp\left(-\beta\sum_{\langle ij \rangle_\alpha}\sum_{\gamma}J^{\alpha\gamma}S_{0,i}^{\gamma
}S_{1,j}^{\gamma} \right)\right]$.\\

It is more convenient to perform the Hubbard-Stratonovich transformation  by representing the Hamiltonian matrix in the basis of the eigenfunctions of the exchange matrix,  which can be easily obtained in the momentum space.
To this end,  we first  introduce a six-component vector ${\bm S}_{\bm q}=(S_{0,{\bm q}}^x,S_{0,{\bm q}}^y,S_{0,{\bm q}}^z,S_{1,{\bm q}}^x,S_{1,{\bm q}}^y,S_{1,{\bm q}}^z)$,  with
the components  given by  the Fourier transforms of the $x,y,z$ components of the spins  on $0-$  and $1-$sublattices, correspondingly.  
This allows  us to write  the Hamiltonian  matrix in the momentum space as
\begin{eqnarray}\label{hammoment}
H=\sum_{\bm q}{\bm S}_{\bm q}^{\dagger }\cdot {\hat J}_{\bm q}\cdot {\bm S}_{\bm q},
\end{eqnarray}%
where  
 the $6\times 6$  exchange matrix ${\hat J}_{\bm q}$ is defined as
\begin{eqnarray}\label{matrixJ}
{\hat J}_{\bm q}=\left(
\begin{array}{cccccc}
0 & 0 & 0 & J_{\bm q}^{x} & 0 & 0 \\
0 & 0 & 0 & 0 & J_{\bm q}^{y} & 0 \\
0 & 0 & 0 & 0 & 0 & J_{\bm q}^{z} \\
\left(J_{\bm q}^{x}\right)^\ast  & 0 & 0 & 0 & 0 & 0 \\
0 & \left(J_{\bm q}^{y}\right)^\ast & 0 & 0 & 0 & 0 \\
0 & 0 & \left(J_{\bm q}^{z}\right)^\ast & 0 & 0 & 0%
\end{array}%
\right),
\end{eqnarray}
with matrix elements given by
\begin{eqnarray}
J_{\bm q}^{\gamma }=\sum_{\alpha =x,y,z}\left[J+K\delta
_{\alpha ,\gamma }\right]
e^{\imath{\bm q}\cdot ({\bm d }_{\alpha }-{\bm d }_{z})}
=J_{{\bm q}}+K_{\bm q}^{\gamma}.
\end{eqnarray}
Here we  drop the overall phase factor $e^{\imath {\bm q}\cdot {\bm d }_{z}}=e^{\imath {\bm q}\cdot (0,1)}=e^{\imath  q_y}$ and denote
$J_{{\bm q}}=J(1+e^{-\imath {\bm q}\cdot{\bm a}_1}+e^{-\imath {\bm q}\cdot{\bm a}_2})$,
 $K_{\bm q}^{\gamma}=K e^{\imath{\bm q}\cdot ({\bm d}_{\gamma }-{\bm d }_{z})}$, where
 $\bm{a}_1\!=\!(\!\frac{\sqrt{3}}{2}\hat{x}+\frac{3}{2}\hat{y}\!) $ and 
$\bm{a}_2\!=\!\sqrt{3}\vec{x}$  are the lattice vectors.
 The matrix ${\hat J}_{\bm q}$ is then diagonalized by a unitary  transformation, ${\hat \kappa}_{\bm q}=U^{-1}_{\bm q}{\hat J}_{\bm q}U_{\bm q}$, leading to the following form of the Hamiltonian
 \begin{eqnarray}
&&H=\sum_{\bm q,\nu} \kappa_{{\bm q},\nu} {\tilde S}^*_{{\bm q},\nu} {\tilde S}_{{\bm q},\nu},
\end{eqnarray}%
where the normal amplitudes of spin-like  variables are defined as
 \begin{eqnarray}
 {\tilde S}^{\nu}_{\bm q} = U_{{\bm q},\nu\eta} S^{\eta}_{\bm q}.
\end{eqnarray}
Note that, depending on the form of the interaction matrix, this transformation in general will mix the spin operators on different sites of the unit cell as well as different components of the spin.
However, in the case of the Kitaev-Heisenberg model, while the two sublattices of the honeycomb lattice are mixed, the $x$, $y$, and $z$ components stay separate.
The partition function (\ref{Z-clas}) then looks like:
\begin{eqnarray}
&&Z=\int\int\left[d\mathbf{S}_{0,j}][d\mathbf{S}_{1,j+{\bm d}_z}\right]\delta(\mathbf{S}_{0,j}^{2}-1)\delta(\mathbf{S}_{1,j+{\bm d}_z}^{2}-1)\nonumber\\&&
\exp\left[-\beta
\sum_{\bm q,\nu} \kappa_{{\bm q},\nu} {\tilde S}^*_{{\bm q},\nu} {\tilde S}_{{\bm q},\nu}\right]
.\label{Z-norm}
\end{eqnarray}
 Following the steps outlined in Refs.\cite{sizyuk15, peter15}, we can separate the mean-field and the fluctuational  contributions to the partition function, $Z=Z_{\rm MF} Z_{\rm fluct}$. In  the Gaussian approximation,  the fluctuation part of the  partition function,
\begin{eqnarray}
Z_{\mathrm{fluct}}=\int[d\varphi
]\exp\left[  -\beta{\mathcal{S}}_{\mathrm{fluct}}\right],
\end{eqnarray}
where
$
{\mathcal{S}}_{\mathrm{fluct}}=\sum_{\mathbf{q};\nu,\nu^{\prime}%
}A_{\mathbf{q,}\nu\nu^{\prime}}\delta\varphi_{\mathbf{q},\nu
}^{\mathrm{\ast}}\delta\varphi_{\mathbf{q},\nu^{\prime}}^{\mathrm{}}
$
 can be obtained by integration over the fluctuation amplitudes $\delta\varphi_{\mathbf{q},\nu}$. 
The explicit expression for the matrix elements of the  fluctuation matrix
${\hat{A}}_{\mathbf{q}}$ computed for an  orientation of the mean-field order parameter  along arbitrary direction $(\sin\theta\cos\phi,\sin\theta\sin\phi,\cos\theta)$ are given in Appendix B.  

Now,  the fluctuation contribution to the  free energy can be written as
\begin{eqnarray}
 {\mathcal F}_{\rm fluct}= -\frac{1}{\beta}
 \ln Z_{\rm fluct}
 =\frac{1}{2\beta}\sum_{\mathbf{q}}
 \ln\left\vert \det\{
 A_{\mathbf{q},\nu\nu^{\prime}}\}\right\vert.
  \end{eqnarray}
  While the mean-field part of the free energy has
the full rotational symmetry, its fluctuational part,  ${\mathcal F}_{\rm fluct}$,
 is sensitive to the direction of the mean-field  order parameter. 
 Thus, by finding the minima  of the fluctuational part of the  free energy,  we can pin the spontaneous
magnetization along some preferred direction of the lattice.

Fig.3 (a)  shows the angular dependence of fluctuational  free energy  ${\mathcal F}_{\rm fluct}(\theta,\phi) $  computed for representative parameters 
$J=-2.9$ meV and $K=8.1$ meV, at which the ground state order is the AF zigzag. The magnitude of
${\mathcal F}_{\rm fluct}(\theta,\phi) $  is presented as a color-coded plot on the
unit sphere, where the minima and maxima of the free energy are shown by the deep
blue and red colors, correspondingly. We see that the minima of ${\mathcal F}_{\rm fluct}(\theta,\phi) $ are achieved when the magnetization is directed
along one of the cubic axes.

This finding shows that the contribution of the fluctuations to the
free energy removes the degeneracy of the ground state found on the mean
field level.
The states which are selected by the thermal fluctuations are the collinear states with the order parameter pointing along one of the cubic axes, thus confirming previous  results of the Monte Carlo simulations\cite{price12,price13,sela14} and spin wave analysis by Chaloupka {et al}.\cite{jackeli10}

\begin{figure}
\includegraphics[width=0.85\columnwidth]{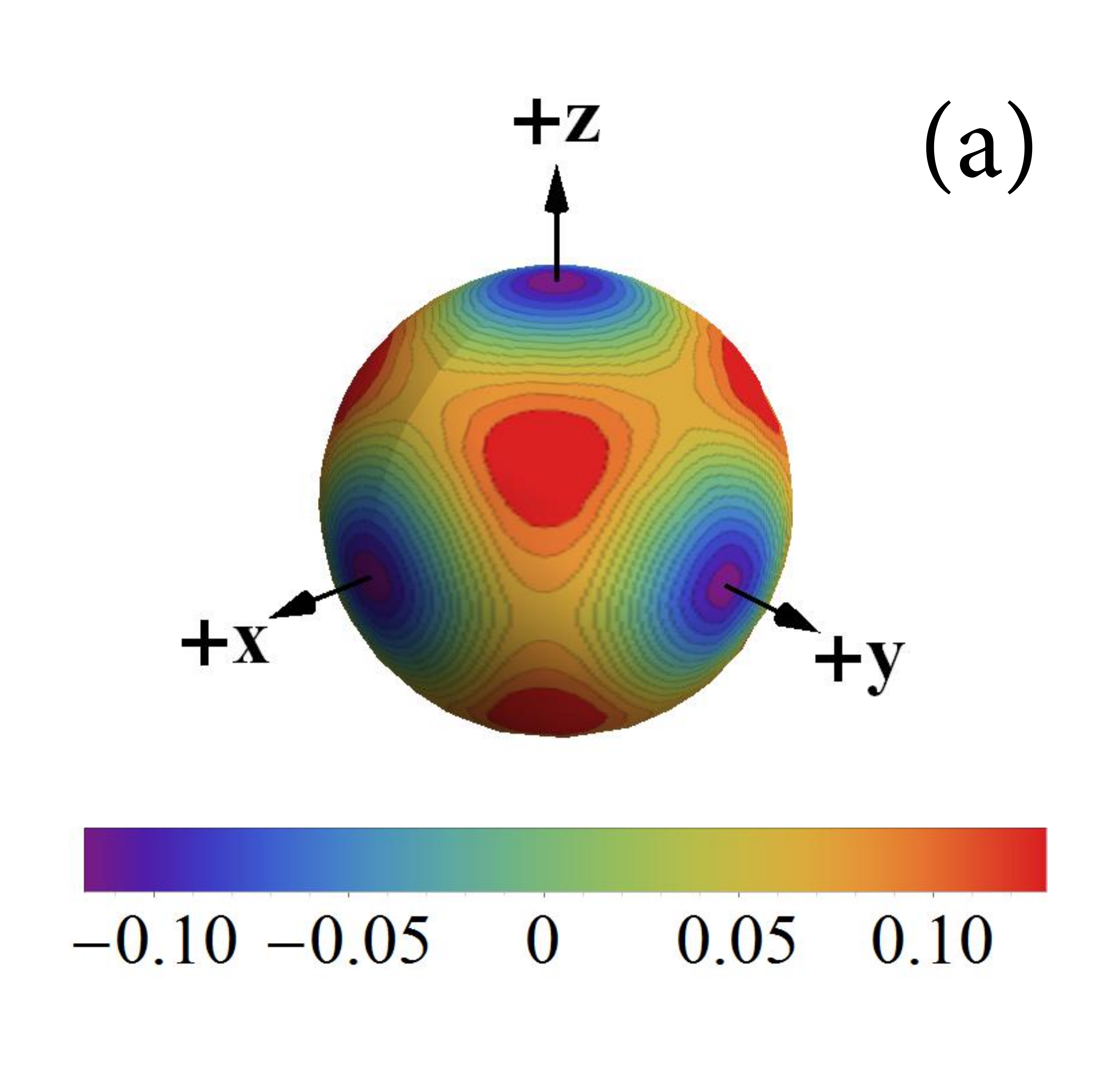}
\includegraphics[width=0.85\columnwidth]{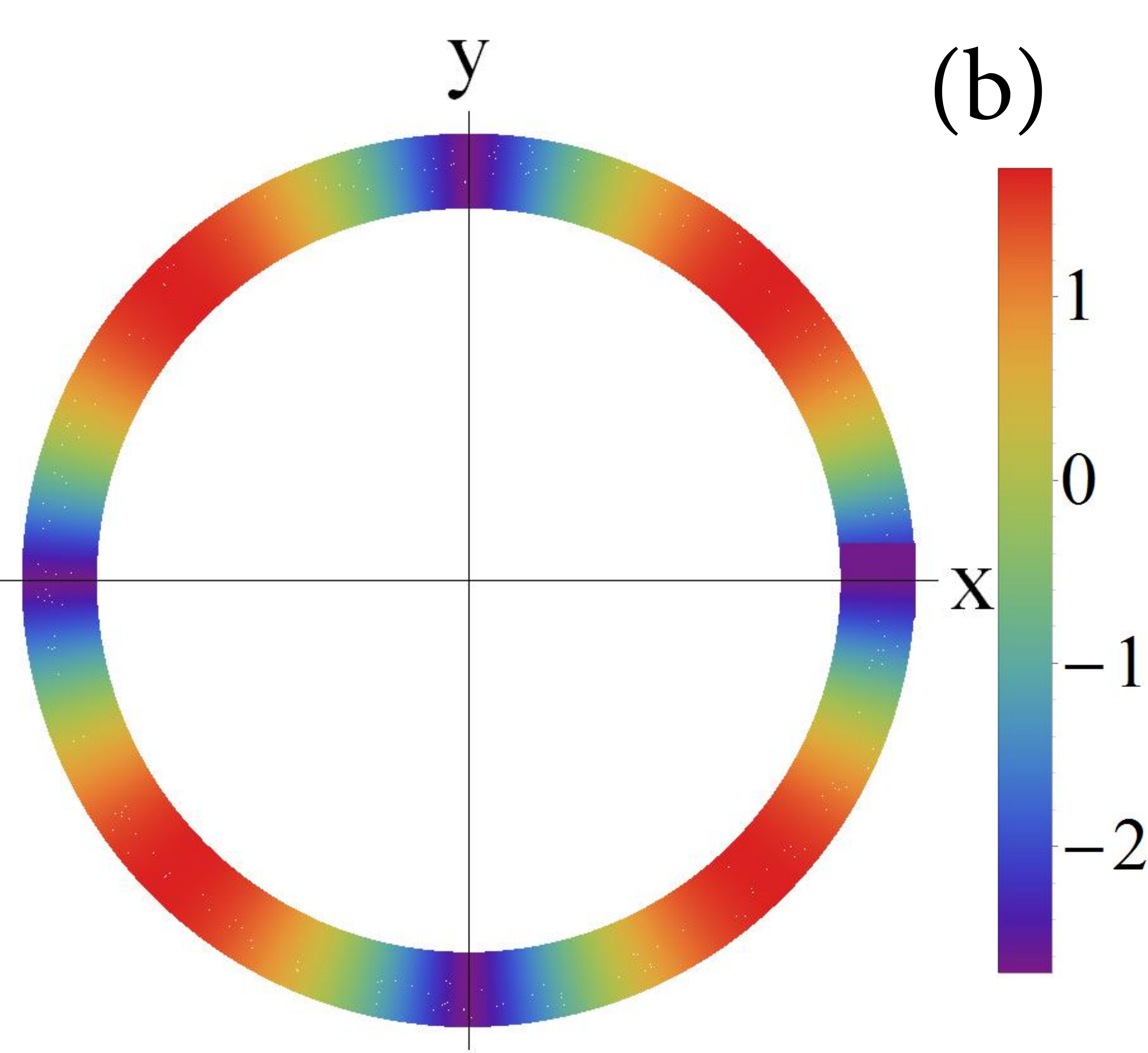}
\caption{
Fluctuational corrections to the free energy in (a) nearest neighbor Kitaev-Heisenberg model computed with  $J=-2.9$ meV and $K=8.1$ meV and (b) $J_1-K_1-J_2-K_2-J_3$ model computed with $J_1=5$ meV, 
$K_1=-17$ meV,  $J_2=-4$ meV,  $K_2=8$ meV, and $J_3=1$ meV. 
}
\label{fig:Nogamma}
\end{figure}

 We discuss the relevance of  our  findings for the nearest neighbor  Kitaev-Heisenberg model for 
 $\alpha-$RuCl$_3$ in Sec. IV.  However in the next section, we will first consider the selection of  the direction of the  order parameter in the extensions of the Kitaev-Heisenberg model relevant for  Na$_2$IrO$_3$.

 \section{Order by disorder in $J_1-K_1-J_2-K_2-J_3$ model}
 
Despite extensive efforts, no consensus concerning the minimal model for Na$_2$IrO$_3$ has
 been reached yet.  The most natural  extension of the  Kitaev-Heisenberg model  with ferromagnetic Kitaev interaction which captures the zigzag magnetic order can be obtained by inclusion of  farther  neighbor couplings. In Na$_2$IrO$_3$, these couplings might not be negligible  due to the extended nature of the $5d$-orbitals of the Ir ions. In the early works suggesting this possible extension,\cite{kimchi11,choi12} second and third neighbor couplings were taken into account phenomenologically and only the isotropic part of these interactions was included.  The importance of  additional nearest neighbor  $C_3$-symmetric anisotropic  terms ($\Gamma$-terms)\cite{rau14,chaloupka15}  or of  the spatial anisotropy of the nearest neighbor Kitaev interactions,\cite{yamaji14}  were also discussed in the literature as a possible source for the stabilization of the zigzag phase.

Here we consider the $J_1-K_1-J_2-K_2-J_3$ model,\cite{sizyuk14}  which 
still has  the same symmetry as the original Kitaev-Heisenberg model  but contains  Kitaev  interactions between both nearest  and second nearest  neighbors.  The model
 reads 
\begin{eqnarray}
&&\mathcal{H}=J_1\sum_{\langle i,j\rangle_{\gamma}}{\bf S}_i{\bf S}_j
+K_1\sum_{\langle i,j\rangle_{\gamma}} S_j^{\gamma}S_j^{\gamma}+
\\\nonumber
&&J_2\sum_{\langle\langle i,j\rangle\rangle_{\tilde\gamma}}
{\bf S}_i{\bf S}_j
+K_2 \sum_{\langle\langle i,j\rangle\rangle_{\tilde\gamma}}S_i^{\tilde\gamma}S_j^{\tilde\gamma}+
J_3\sum_{\langle\langle\langle i,j\rangle\rangle\rangle}
{\bf S}_i{\bf S}_{j},
\end{eqnarray}
where $J_1>0$, $K_1<0$, $J_2<0$, $K_2>0$, and $J_3>0$;
  $\langle\,\rangle$, $\langle\langle \, \rangle\rangle$ and $\langle\langle\langle \, \rangle\rangle\rangle$ denote nearest neighbor, second nearest neighbor and third nearest neighbor, respectively.
 $\gamma=x,y,z$ and $\tilde\gamma={\tilde x}, {\tilde y}, {\tilde z}$ denote the three  types of nearest neighbor and second nearest neighbor bonds of the honeycomb lattice, respectively. It is important  to note   that the second  neighbor  Kitaev interactions 
  do not change the space group symmetries of the original Kitaev-Heisenberg model. 
  
 For realistic sets of the parameters describing Na$_2$IrO$_3$, the second neighbor Kitaev interaction,  $K_2$, computed from the microscopic approach based on the ab-initio calculation by
 Foevtsova {\it et al},\cite{katerina13,note} 
 appeared to  be   the largest interaction after the nearest neighbor Kitaev interaction, $K_1$, and  turn out to be antiferromagnetic.  The
mechanism behind the large magnitude of  $K_2$ in Na$_2$IrO$_3$ is
physically very clear: It originates from the large diffusive Na
ions that reside in the middle of the exchange pathways, and
the constructive interference of a large number of pathways.
Moreover,  the $K_1$-$K_2$ model, that only includes Kitaev interactions,\cite{ioannis15}  already stabilizes  the zigzag phase for the  proper signs of $K_1$ and $K_2$. However, as we have discussed  in Ref.\cite{ioannis15}, the $K_1$-$K_2$ model is still not sufficient to comply with all available experimental data.

The classical degeneracy of the zigzag state obtained within the $J_1-K_1-J_2-K_2-J_3$ model with FM $K_1$ is different from the one  of the zigzag state realized in the extended Kitaev-Heisenberg model with AFM $K_1$ interaction. 
To see what difference the sign of  $K_1$ makes, let us  consider the zigzag pattern in Fig.\ref{fig:orders} (d).
With AFM $K_1$, the pattern, that minimizes the classical energy in the zigzag  state with ferromagnegnetic $y$ and $z$ bonds,  has  the spins  pointing along the $x-$axis to take advantage of the Kitaev interaction on the AFM $x$-bonds.
On the other hand the same pattern with FM $K_1$ takes advantage of the Kitaev interaction on the FM $y-$ and $z-$ bonds by putting spins in the $yz-$plane.
Thus the degenerate ground state manifold for a given zigzag pattern with FM $K_1$ is one of $xy-$, $yz-$, or $zx-$ planes.
Furthermore, when the Klein duality 4-sublattice transformation\cite{jackeli10}
 is applied to the $J_1-K_1-J_2-K_2-J_3$ zigzags, these states do not turn into  N\'{e}el  AFM state, and instead turn into non-collinear states, that are more difficult to work with than the 
original zigzag states. 
Working with the zigzag states directly increases the magnetic unit cell to 4 sites, labeled in Fig \ref{fig:orders}(d).

The Hamiltonian  matrix in the momentum space  can be again written in the form of Eq. (\ref{hammoment}), however this time due to the larger unit cell the exchange matrix ${\hat J}_{\bm q}$ is  $12\times 12$, instead of $6\times 6$.
Its matrix elements are given in Appendix C.
The fluctuations matrix $A_{\mathbf{q,}\nu\nu^{\prime}}$ is calculated as before according to equation (\ref{Amatrix}), with the constraint matrix $C_{{\bm q},\mu,\mu^{\prime}}$ of equation (\ref{Constraint}) now containing 4 identical blocks instead of 2.
The fluctuation matrix again  contains the  information on the direction of the spins and transmits this information to the free energy corrections that we plot in Fig. \ref{fig:Nogamma}(b).
Since the spins are confined to a plane for a given zigzag state we have only the angle of the direction of spins in that plane.
The color of the band at a given angle then gives the size of the fluctuational correction to the free energy, with violet being lowest and red highest energy states.
We see that again the Kitaev anisotropies prefer to align the magnetization along the cubic axes.
Note, however, that unlike the extended KH model, where there were 6 equivalent states, here there are 4 directions for each of the three zigzag patterns, giving a total of 12 states.

\begin{figure}
\includegraphics[width=0.49\columnwidth]{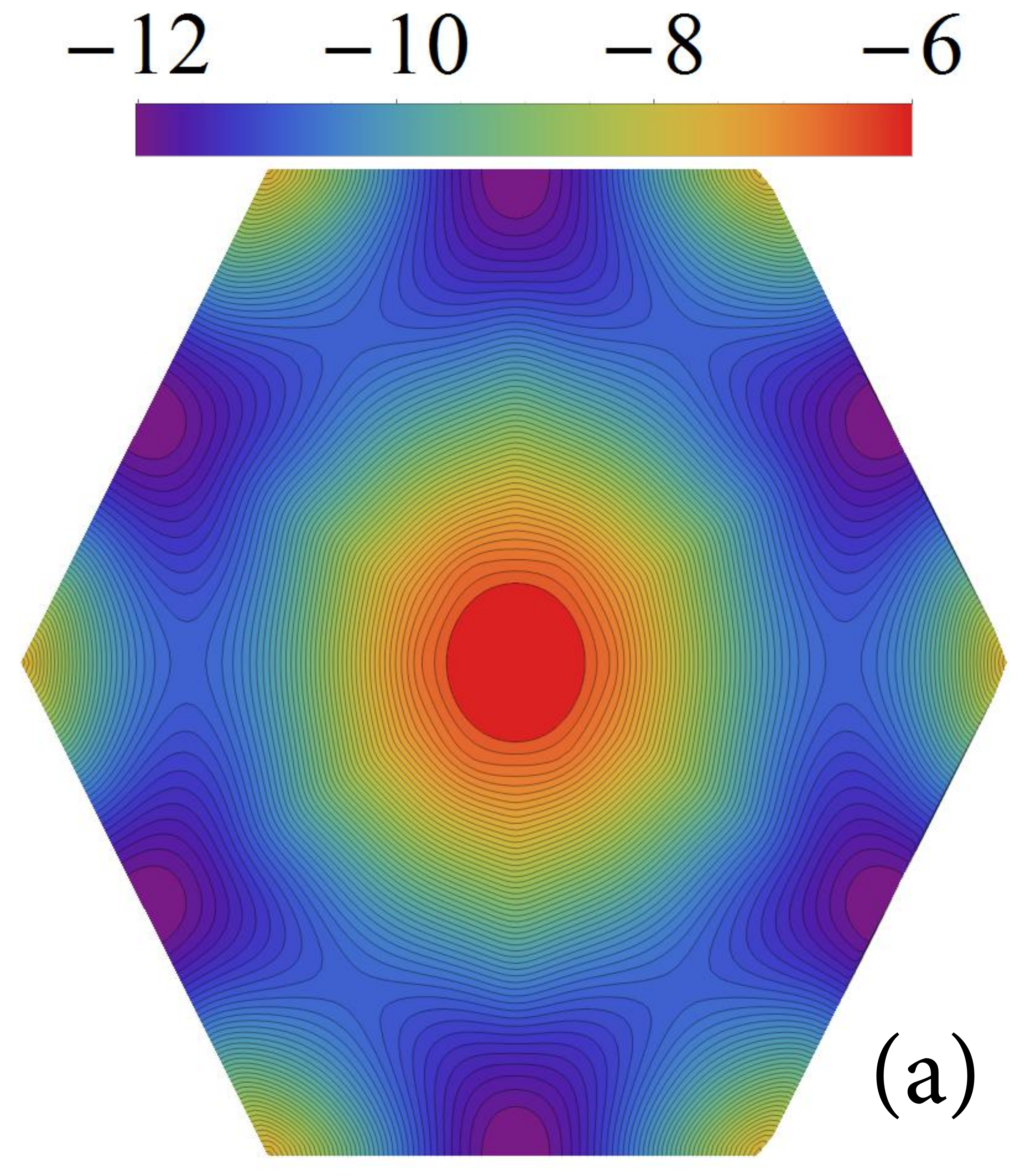}
\includegraphics[width=0.49\columnwidth]{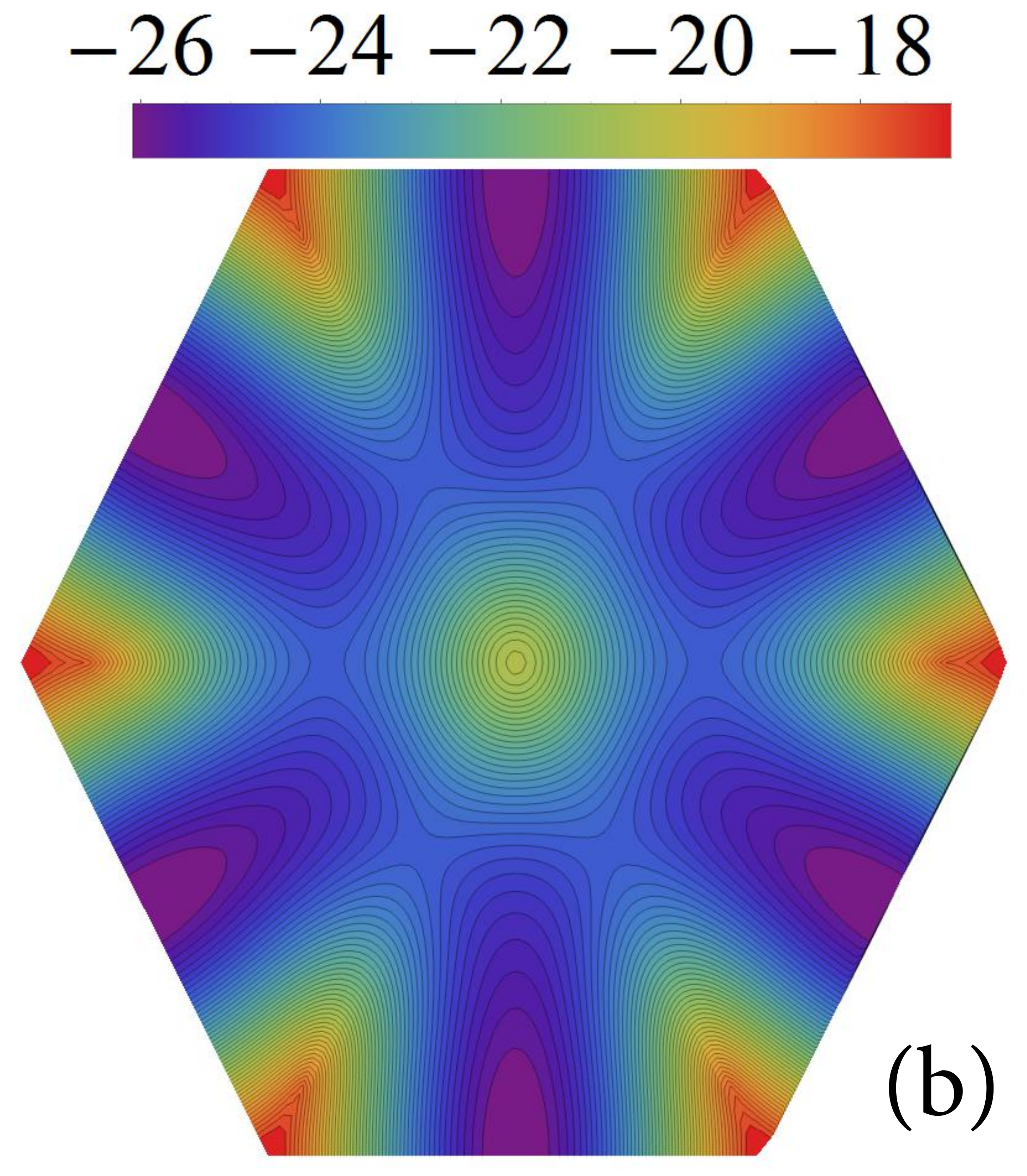}
\includegraphics[width=0.49\columnwidth]{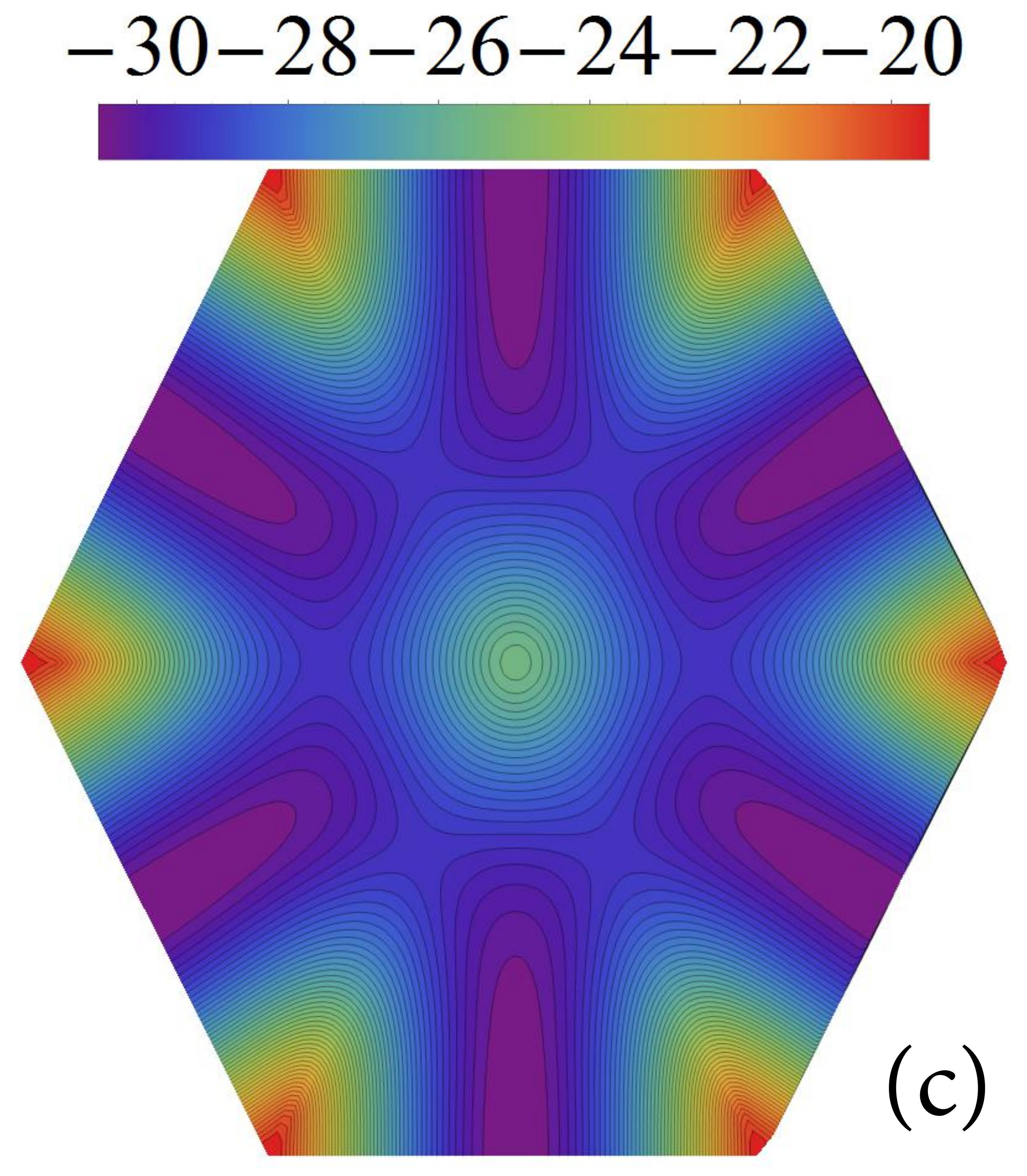}
\includegraphics[width=0.49\columnwidth]{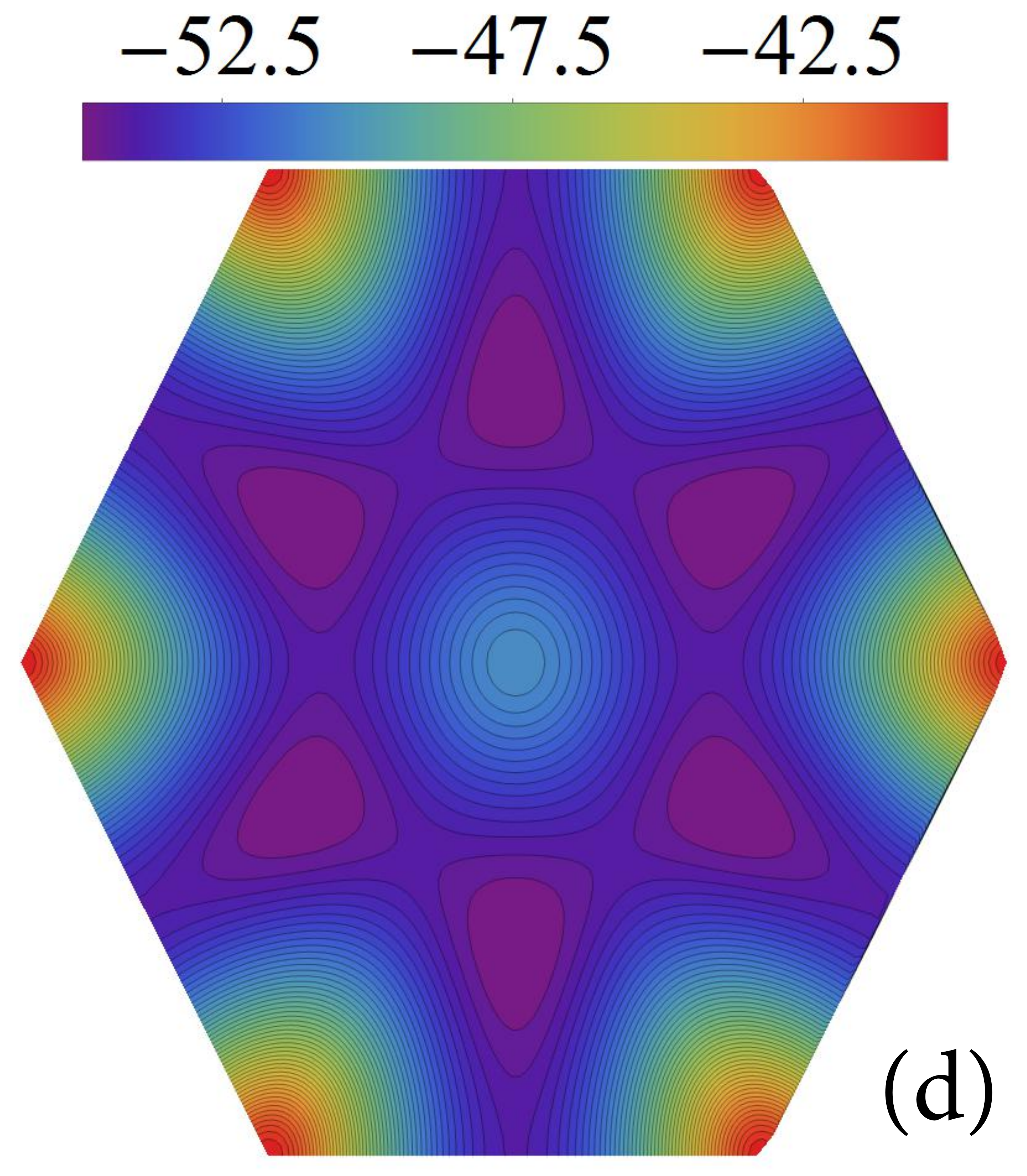}
\caption{
The lowest eigenvalue  of the   $J_1-K_1-J_2-K_2-J_3-\Gamma_1$ model obtained with the Luttinger-Tisza method is shown on the  first Brillouin zone. We use $J_1=3$ meV, $K_1=-17$ meV, $J_2=-3$ meV, $K_2=6$ meV, $J_3=1$ meV, and (a) $\Gamma_1=1$ meV, (b) $\Gamma_1=20$ meV, (c) $\Gamma_1=25$ meV, and (d) $\Gamma_1=50$ meV.
}
\label{fig:LT}
\end{figure}

\section{  The role of  off-diagonal symmetric  $\Gamma$-term. }

\subsection{Directions of the ordered moments in Na$_2$IrO$_3$.} 

 The discussion above has clearly shown, that neither the original Kitaev model nor the extended
  $J_1-K_1-J_2-K_2-J_3$ model can correctly explain the experimental data in Na$_2$IrO$_3$.
Since the easy axes directions are determined solely by the anisotropy terms,  only the inclusion of other types of the anisotropies can improve the situation. Here we consider the off-diagonal symmetric  $\Gamma$-terms.  The role of these terms in the  nearest-neighbor Kitaev model has been studied  in Refs.\cite{rau14,chaloupka15}.  These studies have shown that the  small 
$\Gamma$-terms  do not  immediately  destabilize   the zigzag phase, but lead to a deviation of the magnetic moments from the cubic axes.
 
   The origin of $\Gamma$-terms can be easily seen from the most general form of the bilinear exchange coupling matrix, which on the bond $(i,j)$ has the form given by
\begin{eqnarray} \label{gamma}
\Xi_{i,j}=
\left( \begin{array}{ccc}
J^{xx}\,\,&J^{xy}\,\,&J^{xz}\\
J^{yx}\,\,&J^{yy}\,\,&J^{yz}\\
J^{zx}\,\,&J^{zy}\,\,&J^{zz}\\
\end{array} \right).
\end{eqnarray}
While the Kitaev term comes from the anisotropy of the diagonal matrix elements  of $\Xi_{i,j}$, e.g.  $K_1=J_1^{zz}-J_1^{xx}$,
the symmetric and antisymmetric combinations of off-diagonal elements represent other types of possible bond-anisotropies.   In the absence of the trigonal distortion,  the inversion symmetry prohibits the existence of
 antisymmetric interactions but  some of the symmetric combinations are allowed, i.e. on a given  $\gamma$-bond, the interaction  between the other two spin components, $\Gamma^\gamma (S_i^\alpha S_j^\beta+S_i^\beta S_j^\alpha)$,  where $\Gamma^\gamma= \frac{1}{2}(J_1^{\alpha\beta}+J_1^{\beta\alpha})$,
  is non-zero. Our previous results\cite{sizyuk14} suggest that  in  Na$_2$IrO$_3$  the magnitude of the  strength of   $\Gamma$  on the nearest neighbor bonds is  about   2-3 meV and vanishes for the second neighbors.  
 
 Here we consider the $J_1-K_1-J_2-K_2-J_3-\Gamma$ model with  the previous choice   of Heisenberg and Kitaev interactions   and treat $\Gamma$ as a free parameter. 
A straightforward classical minimization in momentum space using  Luttinger-Tisza approach\cite{Luttinger46,Litvin74,kaplan07}   shows that 
up to very large values of $\Gamma \sim 20$ meV the minima of the classical energy are located at the $M$ points corresponding to the zigzag states.   This is clearly seen in Fig. 4(a)  where we plot the lowest eigenvalues obtained for $\Gamma=1$ meV.
At larger values of $\Gamma$, the minima shift along the lines connecting  $M$ points and  the center of the Brillouin zone (see Fig. 4 (b) for $\Gamma=20$ meV), indicating the  transition to incommensurate order. The incommensurability of the Luttinger-Tisza solution increases  further with larger values of $\Gamma$, which is shown in Figs. 4 (c) and (d).
The exact value of $\Gamma$ at which the transition occurs is difficult to determine due to the transition being so smooth, 
Note, however, that the transition occurs at values of $\Gamma$ far beyond those predicted from our microscopic calculations at ambient pressure.\cite{sizyuk14}

 After we have demonstrated that adding small $\Gamma$ interactions to the $J_1-K_1-J_2-K_2-J_3$ model does not destabilize the zigzag order, let us  now show  that in the presence  of $\Gamma$ the  mean-field degeneracy is  already lifted and
 the preferred directions of the order parameter are selected. This is clearly seen in  Fig. 5 (a) and (b), where the mean field energy of the zigzag order is computed for $\Gamma=1$ meV and $\Gamma=-1$ meV, respectively.
 By inspection, we can see that  non-zero $\Gamma$ selects  the face diagonals  as  easy axes for magnetic ordering, and the sign of $\Gamma$ determines which  of the two face diagonals are preferred.
 For  concreteness, let us consider  the zigzag with AFM $z-$bonds.  As we discussed above the case for  $\Gamma=0$, the easy  $xy$-plane is selected at the mean-field level of  the $J_1-K_1-J_2-K_2-J_3$ model.
 Then, the  inclusion of positive $\Gamma$ interaction on $x$ and $y$ bonds gives zero contribution  to the energy since on these bonds  it involves the spin component perpendicular to the easy plane,
 but it gives  maximal lowering of the energy on the  $z$-bonds if the spins point along $[110]$ and $[{\bar 1}{\bar 1}0]$, $[{\bar 1}10]$ and $[{\bar 1}10]$   directions  correspondingly for positive and negative values of  $\Gamma$. 
 The estimate for the  smallest $\Gamma$, at which the selection of face diagonals takes place, can be done by  comparing the  mean-field  energy gain due to $\Gamma$ with the energy gain due to fluctuations at $\Gamma=0$, which at $T=0$ is equal to the zero point energy and is a function of $K_1$ and $K_2$.   At  finite temperature, the contribution  to the energy from the  Gaussian fluctuations  at each  $T$  can be computed by our method, and this energy will give the lower bound for the magnitude of $\Gamma$ needed to  change the orientation of  magnetic order from  the cubic to the face diagonal.

  \begin{figure}
\includegraphics[width=0.45\columnwidth]{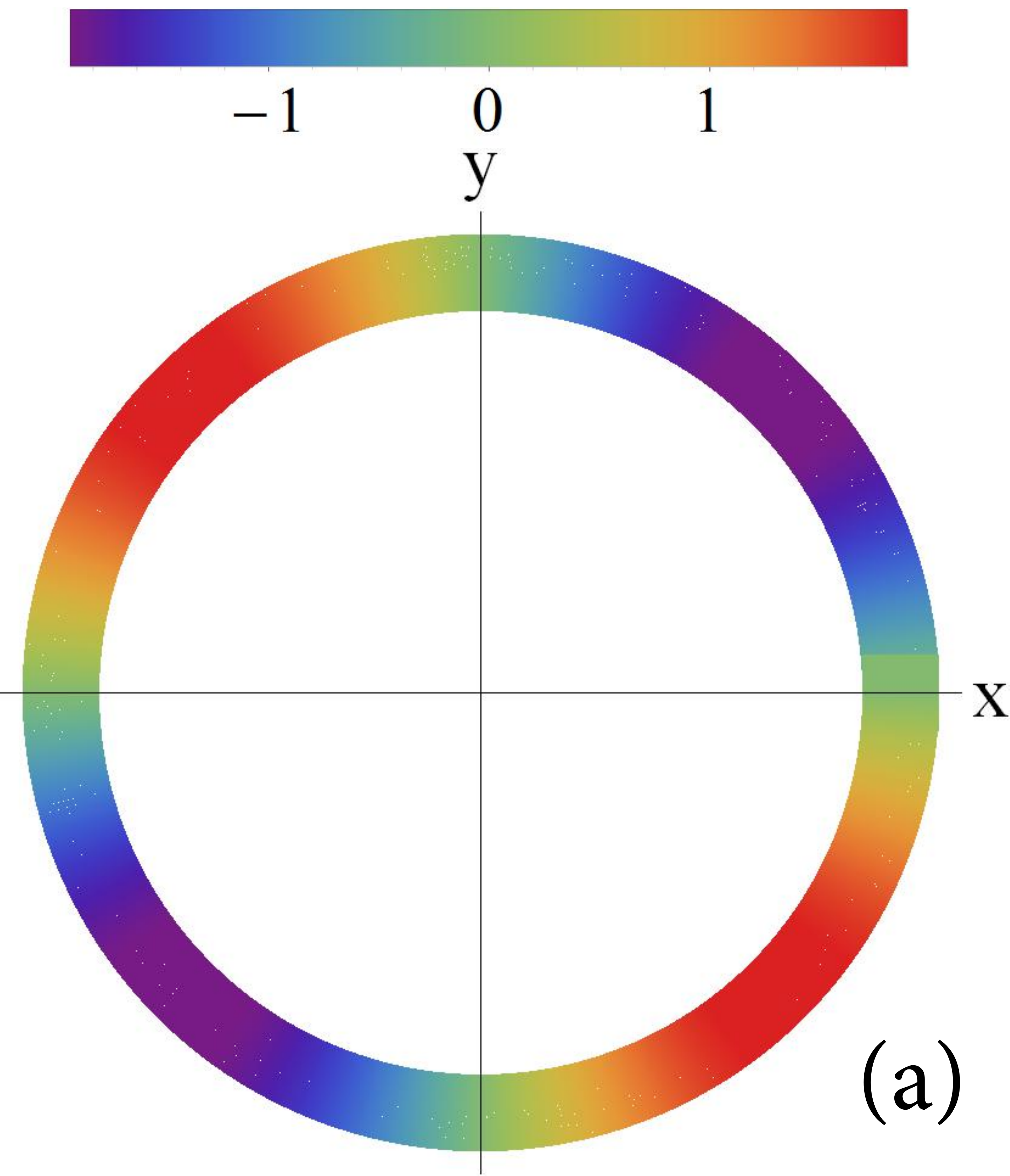}
\includegraphics[width=0.45\columnwidth]{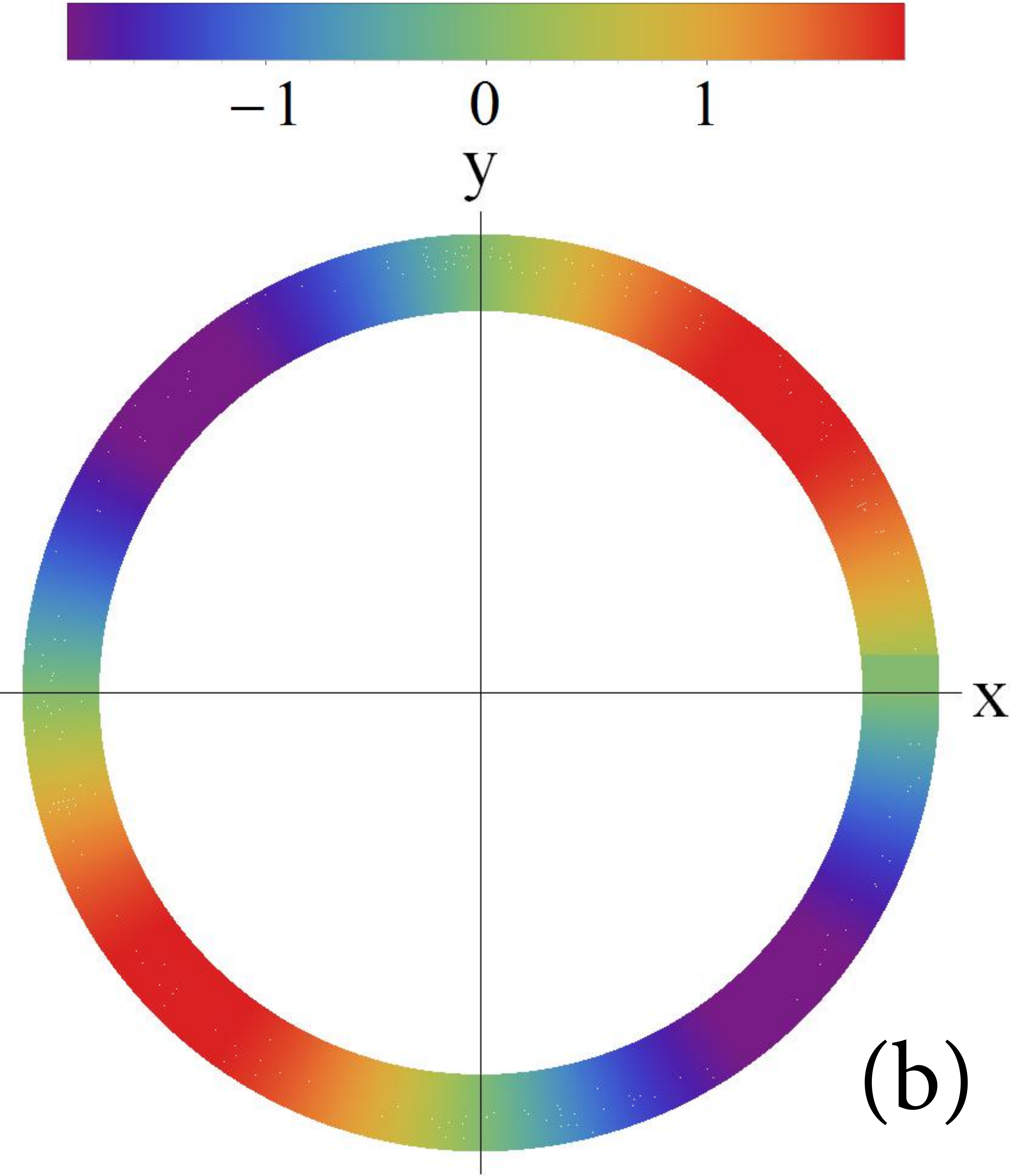}
\caption{
Mean field energy of the zigzag orders in $J_1-K_1-J_2-K_2-J_3$ model with the contribution of (a) $\Gamma=1$ meV and  (b)  $\Gamma=-1$ meV. 
}
\label{fig:Nagamma}
\end{figure}

\subsection{Directions of the ordered moments in $\alpha-$RuCl$_3$.} 

The microscopic calculations for $\alpha-$RuCl$_3$ emphasized the importance of the off-diagonal nearest neighbor  $\Gamma$ interactions.\cite{kee15} The effect of adding $\Gamma$ interaction to the nearest neighbor Kitaev-Heisenberg model is easiest to understand in the rotated reference frame of the four-sublattice Klein transformation.\cite{chaloupka15,kimchi14}
The Kitaev and Heisenberg interactions do not change their form and only change the value of the coupling constants under this transformation. On the other hand, $\Gamma$-interaction picks up a bond dependent sign  as shown in Fig.  2. In fact,  $\Gamma$ changes the sign on half of the bonds, 
 i.e. there are just as many negative bonds as there are positive bonds  for each Kitaev type of bonds.
Since the Klein transformed version of the zigzag state is the AFM N\'{e}el state, all the bonds are AFM and involve the same pair of spins.
Thus the contribution of the $\Gamma$ interaction to the mean-field energy cancels out, and the set of states remains degenerate.
This means that as long as we remain in the small window
where $\Gamma$ does not destabilize the zigzag order found by Rau et al.,\cite{rau14} we can perform our order-by-disorder approach to see what state is chosen. 

Figs.6 (a)-(c)  show the  fluctuation free energy computed for the $J-K-\Gamma$ model  for $J_1=-2.9$ meV and $K_1=8.1$ meV, suggested by Banerjee  {\it et al.},\cite{banerjee15} and $\Gamma=0.7$ meV,  0.8 meV and 0.9 meV, respectively.
In Fig. 6 (a), $\Gamma=0.7$ meV,  the minima of the fluctuational free energy are still along cubic directions.   For larger  $\Gamma$-interaction,  the system prefers the
 states with at least two nonzero spin components and, therefore, the transition towards [111]
  preferred directions of the order parameter takes place. This is shown in  Fig. 6  (b) and (c), in which the fluctuational energy is plotted for 
 $\Gamma=0.8$ meV
  and 0.9 meV.   While in  Fig.  6 (b) only very shallow minima  are seen  along  [111] directions, in
   Fig. 6 (c)  both the pronounced  minima  along  the cubic body diagonals and maxima along  the cubic axes are very clearly seen.  Remember that the computation is done in the rotated reference frame. Therefore, only the states with  the orientation of the order parameter along the cubic axes will give the collinear states in the unrotated reference frame. The  states with order parameter pointing along [111] directions in the rotated reference frame  correspond to non-collinear states in the unrotated reference frame. 
    Since  recent experiments by Cao et al.\cite{cao16} have established that spins point along a cubic axis,  by calculating the fluctuational corrections  as a function of $\Gamma$, we can find an upper bound on its value, such that the Kitaev fluctuations dominate and keep the cubic axes as the preferred directions.  From our calculations it follows that for $J_1=-2.9$ meV and $K_1=8.1$ meV the upper  bound for $\Gamma$ is about 0.8 meV.
Finally,  for this set of parameters the transition to the 120$^{\circ}$- AFM order
occurs around $\Gamma=1.6$ meV.
 Note that this estimate is far smaller than the $\Gamma$ values resulting from {\it ab initio} calculations.\cite{kee15}.

\begin{figure*}
\includegraphics[width=0.65\columnwidth]{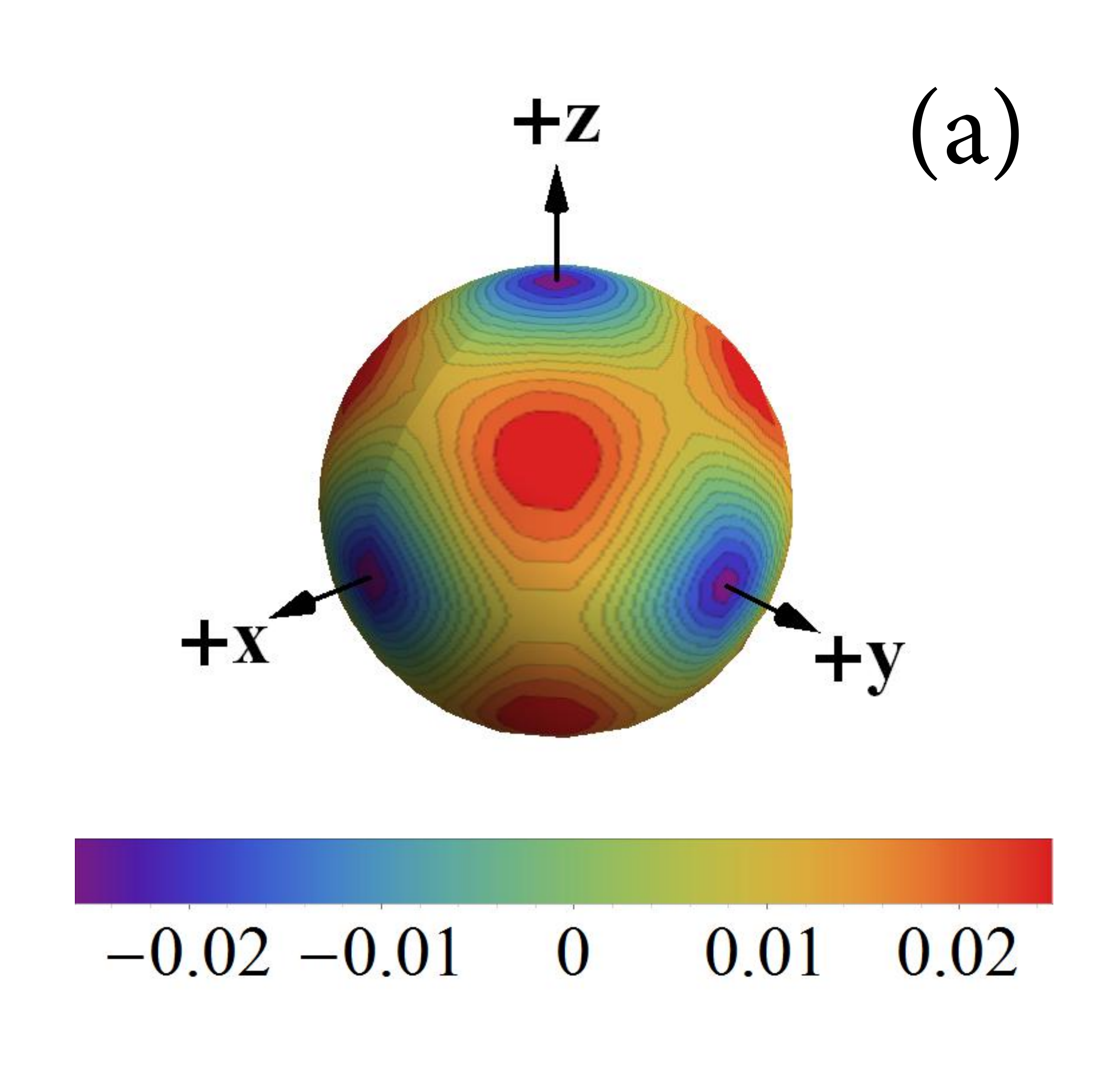}
\includegraphics[width=0.65\columnwidth]{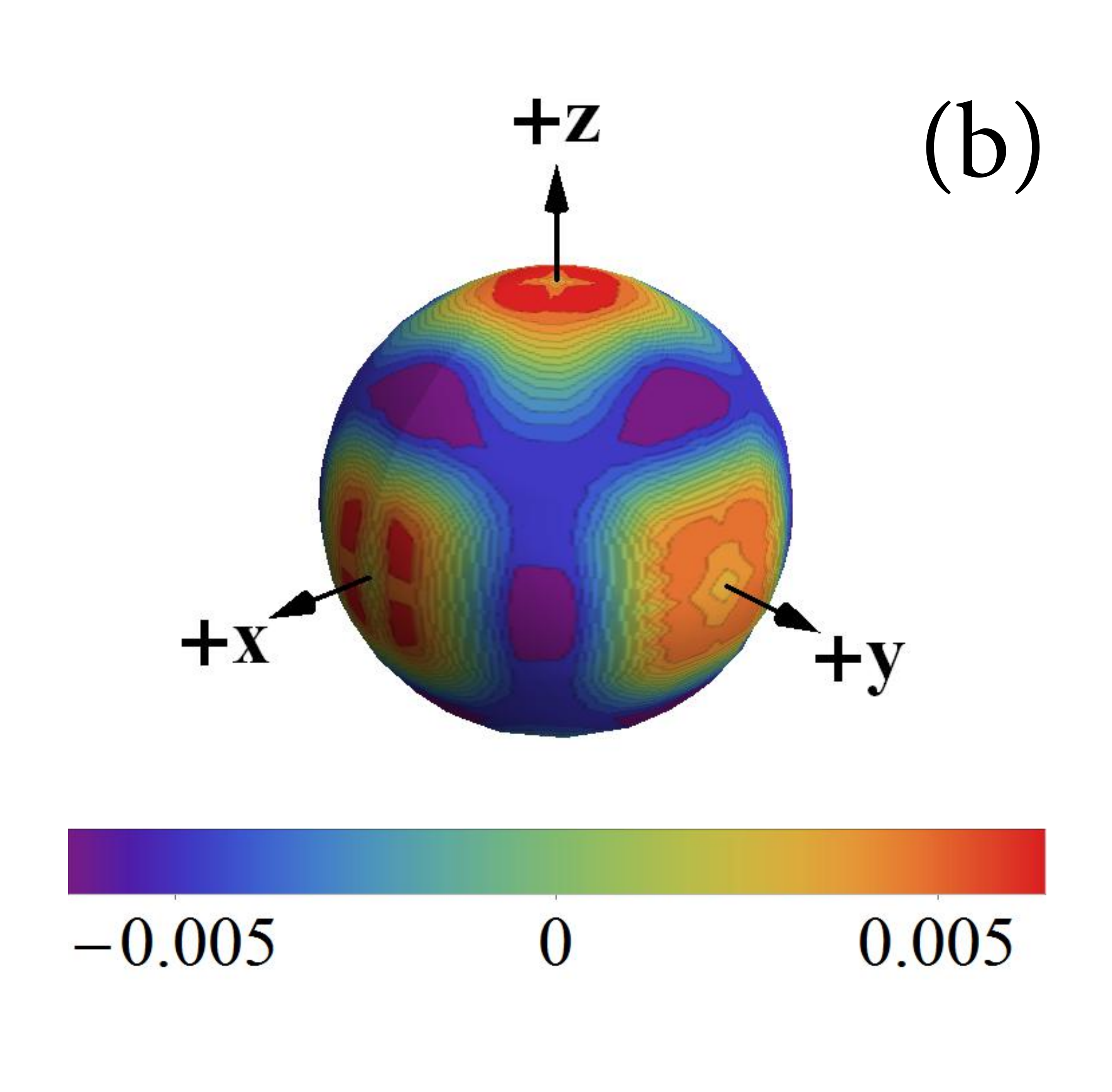}
\includegraphics[width=0.65\columnwidth]{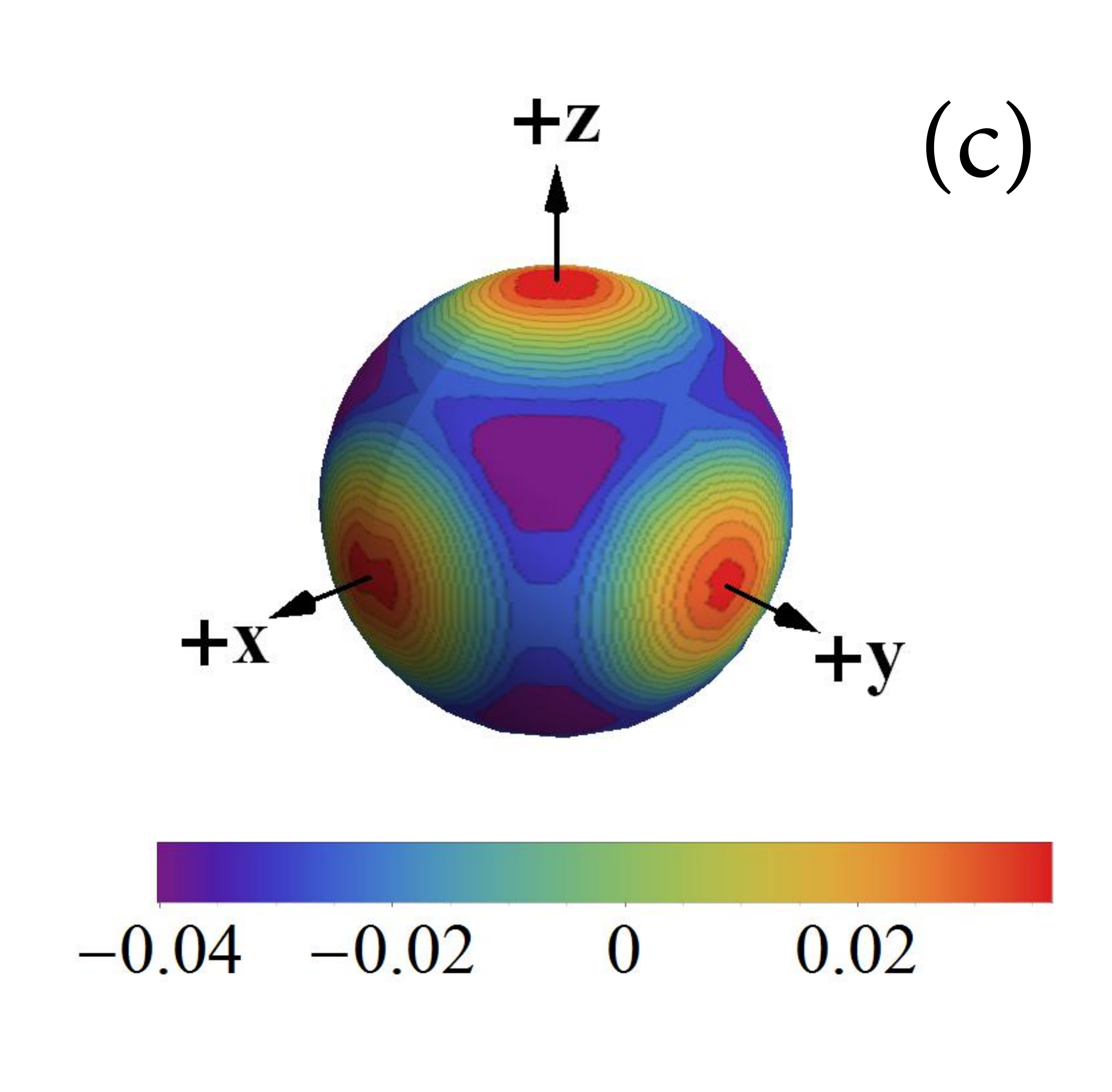}
\caption{ (Color online)
Fluctuational corrections to the free energy in the nearest neighbor Kitaev Heisenberg  model with $\Gamma$ interaction.  We used the following parameters:  $J=-2.9$ meV, $K=8.1$ meV and   
(a) $\Gamma=0.7$ meV, (b) $\Gamma=0.8$ meV, and (c) $\Gamma=0.9$ meV.
The minima of the free energy are shown by deep blue color and the maxima by intense red color.
}
\label{fig:Rugamma}
\end{figure*}

\section{Concluding remarks}

In this paper we  explored how the direction of the magnetic moments in the zigzag ground state order  is chosen in Na$_2$IrO$_3$ and $\alpha-$RuCl$_3$.  
 In both compounds, the Kitaev interaction plays an important role. For  the case of FM nearest neighbor  Kitaev interaction,  like in  Na$_2$IrO$_3$, farther neighbor  interactions are essential for stabilizing the zigzag ground state.  For  the AFM nearest neighbor Kitaev interaction,  which was  widely suggested to be the dominant interaction in   $\alpha-$RuCl$_3$,\cite{sears15,johnson15,banerjee15,cao16,kee15} the zigzag order can be stabilized already within the nearest neighbor model.
 
 We proposed that the $J_1-K_1-J_2-K_2-J_3-\Gamma$ model can explain all the experimental finding in Na$_2$IrO$_3$.   In this model the selection of the experimentally observed face  diagonal direction of the order parameter  happens  already on the mean-field level
 due to the small bond-dependent anisotropic term $\Gamma$. 

 In $\alpha-$RuCl$_3$,  if the the nearest neighbor  Kitaev interaction is AFM, the original Kitaev-Heisenberg model\cite{jackeli13} is sufficient to explain both the collinear zigzag  ground state and the cubic directions of the order parameter.  We  studied the effect   of the $\Gamma$-term  and showed that while  on the mean-field level it  doesn't affect  the ground state degeneracy, 
 it   favors non-collinear 3-{\bf Q} states, instead of the experimentally observed zigzag state with spins along cubic axes, once the Gaussian fluctuations are included. Thus, it appears to be 
an upper bound for $\Gamma$-term, which can be  estimated for a given set of   nearest  neighbor parameters.

After  the  completion  of  our  paper,  we became aware of an independent study by   Winter {\it  et al}.
\cite{roser16}  of the magnetic interactions in the Kitaev materials  Na$_2$IrO$_3$ and $\alpha-$RuCl$_3$. In this  work, the authors treated all interactions up to third neighbours on equal footing  by 
combining  exact diagonalization and ab-initio  techniques.   One of the main findings of this work is that the third neighbor Heisenberg  interaction   is important in all Kitaev materials. 

Let us briefly compare the results of  Ref. \cite{roser16} with our findings.
The conclusions of the authors of  Ref. \cite{roser16} about the ordering in Na$_2$IrO$_3$  are in agreement with our findings,  despite the fact  that their estimates for $K_2$  suggest significantly smaller values  than  the ones that we obtained by  including only the dominant  superexchange processes between the second neighbors. The agreement holds  because the second neighbor Kitaev interaction $K_2$ and the third neighbor interaction $J_3$   favor the same type of AFM zigzag  ground state. 

 For $\alpha-$RuCl$_3$, the  authors of Ref.\cite{roser16}  suggest  ({\it i})  that there may be possible
 variations  of in-plane interactions   due to lattice distortions, and ({\it ii})  that the  nearest neighbor Kitaev interaction  may be FM and the third neighbor coupling $J_3$ may be large and AFM.
 The FM sign  of the  nearest neighbor Kitaev interaction was also suggested   by  Yadav {\it et al }  in Ref.\cite{hozoi16}.
  If this is indeed the case, the physics of  $\alpha-$RuCl$_3$ is  similar to   that of  Na$_2$IrO$_3$.
  This, however,   still  needs to be verified by a detailed comparison  with the experimental data.

{\it Acknowledgements.}
We acknowledge insightful discussions with  C. Batista,   G. Jackeli, M.Garst, G. Khaliullin and I. Rousochatzakis.
N.P. and Y.S. acknowledge the support from NSF Grant  DMR-1511768.
N.P. acknowledges the hospitality of KITP and partial support
 by the National Science Foundation under Grant No. NSF PHY11-25915.

\appendix

\section{ The classical degeneracy of the extended Kitaev-Heisenberg model}

 In this Appendix we provide detailed discussion of  the classical degeneracy of the extended Kitaev-Heisenberg model at parameters for which  either the stripy or the zigzag AF phases are realized as the ground state and the  manifold of classically 
degenerate states  is rather complex. 

To be specific, let us first consider the stripy phase. It contains six inequivalent collinear stripy states with FM bonds along  either Kitaev $x-$, $y-$ or $z-$bonds.
It also contains infinite number of non-collinear (coplanar and non-coplanar) states.  The spin  order in the  $x-$, $y-$ or $z-$ stripy states can be described either with a help  of four magnetic sublattices or 
 by  a simple spiral  characterized by a single-$Q$  wave vector: ${\mathbf Q}_x=(\pi/\sqrt{3},\pi)$, ${\mathbf Q}_y=(\pi/\sqrt{3},-\pi)$ and ${\mathbf Q}_z=(-2\pi/\sqrt{3},0)$. 
 One of the stripy  states with FM $z$-bonds  is shown in Fig.1 (c).  
In each of these stripy  states the spins are aligned along  one of  the cubic directions which is locked to the spatial orientation of a stripy pattern by the Kitaev interaction, 
i.e.  the direction of the order parameter is defined by the wave vector ${\mathbf Q}={\mathbf Q}_x, {\mathbf Q}_y$ or ${\mathbf Q}_z$ determining the breaking of the translation symmetry.

 The structure of the manifold of the non-collinear states, which looks  rather complex in the original model,  can be easily understood  with the help of the four-sublattice  transformation (see Fig.2) based on the Klein duality.\cite{jackeli10,chaloupka15,kimchi14}
In the new  rotated basis, the stripy phase is mapped to the FM order with a unique ordering vector ${\mathbf Q}=0$. Classically, all states with arbitrary direction of the FM order have the same energy. 
FM states with order parameter along the cubic axes give the six stripy phases in the unrotated spin basis discussed above. 
Arbitrary  directions of the FM order parameter lead to a set of non-coplanar states  in which each component of spin, $S_x$, $S_y$, and $S_z$, transforms  with its own  ${\mathbf Q}_x$, ${\mathbf Q}_y$ and ${\mathbf Q}_z$  wavevector,
which coincide with the ${\mathbf Q}$ vectors describing  the spatial orientation of the stripes in the respective collinear states. 

Using these three ordering vectors, we can write the non-coplanar phase of the unrotated spins as
\begin{eqnarray}\label{0}
{\mathbf S}_{i,0}=(s_\theta c_\phi \,e^{\imath{\mathbf Q}_x\cdot{\bm R}_i},s_\theta s_\phi\, e^{\imath{\mathbf Q}_y\cdot{\bm R}_i},c_\theta \,e^{\imath{\mathbf Q}_z\cdot{\bm R}_i}),
\end{eqnarray}
where $\theta$ and $\phi$ are the polar and azimuthal angles of the FM order parameter. ${\mathbf S}_{i,0}$ denote  the spins on the sublattice $0$ and 
the spins on the sublattice $1$ are obtained  from ${\mathbf S}_{i,0}$  by a constant phase shift coming from the spin rotation on that bond as prescribed by the four sublattice transformation.
As in Fig.1 (c),  the sublattices 0 and 1 are connected by the $z$ bond, the order of the spins on the subllatice  1 is given by
\begin{eqnarray}\label{1}
{\mathbf S}_{i,1}=(S_{i,0}^x\, e^{\imath\pi},\,S_{i,0}^y\, e^{\imath \pi},S_{i,0}^z)
\end{eqnarray}

In the zigzag phase, the structure of the classical states manifold is very similar to the stripy phase.
 The four-sublattice transformation maps the zigzag phase onto the N\'{e}el   AF phase. The generic state is again described by the three-${\mathbf Q}$ spiral state. The only difference is that the spins on sublattice 1 have an overall phase factor of $\pi$,
 ${\mathbf S}_{i,1}=(S_{i,0}^x,\,S_{i,0}^y,\,S_{i,0}^ze^{\imath\pi})$.

\section{The  matrix elements  $A_{\mathbf{q,}\nu\nu^{\prime}}$  computed for the KH model. }

The matrix elements $A_{\mathbf{q,}\nu\nu^{\prime}}$  can be written as
\begin{eqnarray}\label{Amatrix}
A_{\mathbf{q,}\nu\nu^{\prime}}=\frac{\delta_{\nu, \nu^\prime}}{\kappa_{{\bm q},\nu}}+
s(\kappa_{{\bm q},\nu})s(\kappa_{{\bm q},\nu^\prime}) U^{-1}_{{\bm q},\nu,\mu}C_{{\bm q},\mu,\mu^{\prime}} U_{{\bm q},\nu,\mu},
\end{eqnarray}
where a repeated index implies a summation over.
 The first term in (\ref{Amatrix}) is the contribution from the interaction term and the second term is from the constraint term.\cite{sizyuk15, peter15}
For convenience, the constraint matrix  ${\hat C}_{{\bm q}} $  can be  first written in the original basis, in which the interaction term is not diagonal, and then transformed to the eigenbasis  of the Hamiltonian with a help of the
unitary transformation $U_{\bm q}$.
In the original basis the constraint matrix ${\hat C}_{{\bm q}} $ consists of two blocks, one for 
each sublattice. The A-sublattice block has  elements  $C_{{\bm q},\mu,\mu^{\prime}} $ with $\mu,\mu^{\prime}=1,2,3$ and the B-sublattice  block has the elements  with $\mu,\mu^{\prime}=4,5,6$.
The two blocks are  identical,  so  ${\hat C}_{{\bm q}} $ takes the following form:
\begin{eqnarray}
{\hat C}_{\bm q}=\left(
\begin{array}{cccccc}
C_{{\bm q},11}&C_{{\bm q},12}&C_{{\bm q},13}& 0 & 0 & 0  \\
C_{{\bm q},21}&C_{{\bm q},22}&C_{{\bm q},23}&0 & 0 & 0  \\
C_{{\bm q},31}&C_{{\bm q},32}&C_{{\bm q},33}&0 & 0 & 0 \\
 0 & 0 & 0&C_{{\bm q},11}&C_{{\bm q},12}&C_{{\bm q},13}  \\
0 & 0 & 0  &C_{{\bm q},21}&C_{{\bm q},22}&C_{{\bm q},23}\\
0 & 0 & 0 &C_{{\bm q},31}&C_{{\bm q},32}&C_{{\bm q},33}
\end{array}%
\right)
\end{eqnarray}
with matrix elements given by
\begin{equation}\label{Constraint}
\begin{array}
[c]{l}%
C_{{\bm q},11}=
-\frac{2}{3}\bigl[
\beta_c(1-s_{\theta}^{2}c_{\phi}^{2})+3\beta r s_{\theta}^{2}c_{\phi}^{2}\bigr],
\\
C_{{\bm q},22}=
-\frac{2}{3}\bigl[
\beta_c(1-s_{\theta}^{2}s_{\phi}^{2})+3\beta r s_{\theta}^{2}s_{\phi}^{2}\bigr],
\\
C_{{\bm q},33}=
-\frac{2}{3}\bigl[
\beta_c s^2_{\theta}+3\beta r c^2_{\theta}\bigr],
\\
C_{{\bm q},12}=C_{{\bm q},21}=-\frac{2}{3}
(3\beta r-\beta_c) s^2_{\theta} c_{\phi}s_{\phi},
\\
C_{{\bm q},13}=C_{{\bm q},31}=-\frac{2}{3}
(3\beta r-\beta_c) s_{\theta}s_{\theta} c_{\phi},
\\
C_{{\bm q},23}=C_{{\bm q},32}=-\frac{2}{3}
(3\beta r-\beta_c) s_{\theta}s_{\phi} c_{\phi},
\end{array}
\label{Cmatrix}%
\end{equation}
where, to shorten notations, we denote $\sin\theta(\phi)\equiv s_{\theta
(\phi)}$ and $\cos\theta(\phi)\equiv c_{\theta(\phi)}$.

\section{  Coupling $J_{\mu,\nu}({\bm q})$ of the $J_1-K_1-J_2-K_2-J_3$ model. }

For shortness we define 
$q_1={\bf q}\cdot{\bf a}_1$, 
$q_2={\bf q}\cdot{\bf a}_2$, 
and $q_z={\bf q}\cdot{\bf d}_z$.
The  diagonal matrix elements for $\mu=1,4,7$ and 10 are equal to  
$J_{\mu,\mu}({\bm q})=(J_2+K_2)\cos  q_1$, all other diagonal elements are equal to
$J_{\mu,\mu}({\bm q})= 
J_2\cos  q_1$.
The non-zero off-diagonal elements $J_{\mu,\nu}({\bm q})$  for  $\nu > \mu$ are
\begin{eqnarray}
&&J_{1,4}({\bm q})= \frac{1}{2}
J_1 (e^{ \imath q_z}+e^{ \imath(-q_1+ q_z)})
\nonumber\\
&&J_{2,5}({\bm q})= \frac{1}{2}
\left(J_1 (e^{ \imath q_z}+(J_1+K_1) e^{ \imath(-q_1+q_z)})\right)
\nonumber\\
&&J_{3,6}({\bm q})= \frac{1}{2}
\left((J_1+K_1) (e^{ \imath q_z}+J_1 e^{ \imath(- q_1+q_z)})\right)
\nonumber\\
&&J_{1,7}({\bm q})= 
J_2 ( \cos (q_1-q_2)+ \cos  q_2
)
\nonumber\\
&&J_{2,8}({\bm q})= 
(J_2+K_2 ) \cos(q_1-q_2)+ J_2\cos q_2
\nonumber\\
&&J_{3,9}({\bm q})= 
J_2  \cos(q_1-q_2)+ (J_2+K_2 ) \cos q_2\nonumber
\\
&&J_{1,10}({\bm q})= \frac{1}{2}
\Big(
(J_1 +K_1)e^{ \imath(q_2-q_1+q_z)} +
 \, \nonumber\\&&
\left.\,\,\,\,\,\,\,\,\,\,\,\,\,
J_3 (
e^{ \imath(q_2+q_z)}+e^{ \imath(q_2-2q_1+q_z)}+
e^{ \imath(-q_2+q_z)}
)\right)
\nonumber\\
&&J_{2,11}({\bm q})= \frac{1}{2}\Big(
J_1e^{ \imath(q_2-q_1+q_z)}+\nonumber\\&&
\left.\,\,\,\,\,\,\,\,\,\,\,\,\,
J_3 (
e^{ \imath(q_2+q_z)}+e^{ \imath(q_2-2q_1+q_z)}+
e^{ \imath(-q_2+q_z)}
)\right)
\nonumber\\
&&J_{3,12}({\bm q})= \frac{1}{2}\Big(
J_1e^{ \imath(q_2-q_1+q_z)}+\nonumber\\&&
\left.\,\,\,\,\,\,\,\,\,\,\,\,\,
J_3 (
e^{ \imath(q_2+q_z)}+e^{ \imath(q_2-2q_1+q_z)}+
e^{ \imath(-q_2+q_z)}
)\right)
\nonumber\\
&&J_{4,7}({\bm q})= \frac{1}{2}\left(
(J_1 +K_1)e^{ \imath (q_1-q_2-q_z)}+\right.\nonumber\\&&
\left.\,\,\,\,\,\,\,\,\,\,\,\,\,
J_3 (
e^{ \imath(2q_1-q_2-q_z)}
+e^{ \imath(-q_2-q_z)}
+e^{ \imath(q_2-q_z)}
)\right)
\nonumber\\
&&J_{5,8}({\bm q})= \frac{1}{2}\left(
J_1e^{ \imath (q_1-q_2-q_z)}+\right.\nonumber\\&&
\left.\,\,\,\,\,\,\,\,\,\,\,\,\,
J_3 (
e^{ \imath(2q_1-q_2-q_z)}
+e^{ \imath(-q_2-q_z)}
+e^{ \imath(q_2-q_z)}
)\right)
\nonumber\\
&&J_{6,9}({\bm q})= \frac{1}{2}\left(
J_1e^{ \imath (q_1-q_2-q_z)}+\right.\nonumber\\&&
\left.\,\,\,\,\,\,\,\,\,\,\,\,\,
J_3 (
e^{ \imath(2q_1-q_2-q_z)}
+e^{ \imath(-q_2-q_z)}
+e^{ \imath(q_2-q_z)}
)\right)
\nonumber\\
&&J_{4,10}({\bm q})= 
J_2 \Big( \cos q_2+ \cos (q_2-q_1)
\Big)
\nonumber\\
&&J_{5,11}({\bm q})= 
J_2 \cos q_2+(J_2+K_2) \cos (q_2-q_1)
\nonumber\\
&&J_{6,12}({\bm q})= 
(J_2+K_2) \cos q_2+J_2 \cos (q_2-q_1)
\nonumber\\
&&J_{7,10}({\bm q})= \frac{1}{2}
J_1 (
e^{ \imath q_z}+
e^{ \imath(-q_1+q_z)})
\nonumber\\
&&J_{8,11}({\bm q})= \frac{1}{2}\Big(
J_1 
e^{ \imath q_z}+
(J_1+K_1)e^{ \imath(-q_1+q_z)}\Big)
\nonumber\\
&&J_{9,12}({\bm q})= \frac{1}{2}\Big(
(J_1 +K_1)
e^{ \imath q_z}+J_1
e^{ \imath(-q_1+q_z)}\Big)
\nonumber
\end{eqnarray}


\begin{thebibliography}{100}
\bibitem{note}
   In Ref. \cite{sizyuk14} we obtained large $K_2$ interaction by considering only dominant  super-exchange processes between second neighbors.  The authors of  recent  study\cite{roser16} claim  that the second neighbor Kitaev interaction might be  suppressed due to the interference of the various second
and third order hopping processes, which  we did not include  in our derivation. However, as it is discussed in the text, the combined effect of $K_2$ and $J_3$ interactions leads to the same  physics.
\end{thebibliography}
\end{document}